\documentclass{article}

\usepackage{arxiv}

\usepackage[utf8]{inputenc} 
\usepackage[T1]{fontenc}    
\usepackage{hyperref}       
\usepackage{url}            
\usepackage{booktabs}       
\usepackage{amsfonts, amsmath}       
\usepackage{nicefrac}       
\usepackage{microtype}      
\usepackage{lipsum}		
\usepackage{graphicx}
\usepackage{natbib}
\usepackage{doi}

\usepackage{comment}

\usepackage{xcolor}
\hypersetup{
	colorlinks,
	linkcolor={red!50!black},
	citecolor={blue!50!black},
	urlcolor={blue!80!black}
}

\usepackage[caption=false]{subfig}
\usepackage{sidecap}

\usepackage{algorithm}

\usepackage{braket,amsfonts}

\usepackage{array}

\usepackage[caption=false]{subfig}
\usepackage{sidecap}

\usepackage{pgfplots}

\usepackage{afterpage}
\usepackage{tikz}
\usetikzlibrary{patterns}
\usepackage{array,multirow}

\renewcommand{\vec}[1]{\boldsymbol{#1}}
\renewcommand{\th}{\boldsymbol{\theta}}
\newcommand{\x}{\vec{x}}
\newcommand{\y}{\boldsymbol{y}}
\newcommand{\Q}{\boldsymbol{Q}}
\renewcommand{\L}{\boldsymbol{L}}

\renewcommand{\S}{\boldsymbol{\Sigma}}

\definecolor{mybeige1}{HTML}{FFEBCD} 
\definecolor{mybeige2}{HTML}{FAE7B5}

\definecolor{myblue}{RGB}{30,125,186}
\definecolor{myred}{RGB}{192, 50, 33}

\definecolor{armygreen}{rgb}{0.29, 0.33, 0.13}

\definecolor{asparagus}{rgb}{0.53, 0.66, 0.42}
\definecolor{bittersweet}{rgb}{1.0, 0.44, 0.37}
\definecolor{armygreen}{rgb}{0.29, 0.33, 0.13}
\definecolor{antiquefuchsia}{rgb}{0.57, 0.36, 0.51}

\definecolor{myorange}{HTML}{FB9800}
\definecolor{myorange2}{HTML}{FAC898}
\definecolor{myyellow}{HTML}{FFCC80}

\usepackage{algorithmic}

\title{Integrated nested Laplace approximations for \\ large-scale spatial-temporal Bayesian modeling}

\author{Lisa Gaedke-Merzhäuser \\
	Faculty of Informatics\\
	Università della Svizzer italiana\\
	Lugano, Switzerland \\
	\textit{lisa.gaedke.merzhaeuser@usi.ch} \\
	\And
	Elias Krainski \\
	CEMSE Division\\
	KAUST \\
	Thuwal, Saudi Arabia \\
	\textit{elias.krainski@kaust.edu.sa} \\
	\AND
	Radim Janalik \\
	Faculty of Informatics\\
	Università della Svizzer italiana\\
	Lugano, Switzerland \\
	\textit{radim.janalik@usi.ch} \\
	\And
	Håvard Rue \\ 
	CEMSE Division \\
	KAUST \\
	Thuwal, Saudi Arabia \\
	\textit{haavard.rue@kaust.edu.sa}
	\And 
	Olaf Schenk \\
	Faculty of Informatics\\
	Università della Svizzer italiana\\
	Lugano, Switzerland \\
	\textit{olaf.schenk@usi.ch} \\
}

\renewcommand{\shorttitle}{Large-Scale Spatial-Temporal Bayesian modeling using INLA}

\begin{document}
\maketitle

\begin{abstract}
Bayesian inference tasks continue to pose a computational challenge. This especially holds for spatial-temporal modeling where high-dimensional latent parameter spaces are ubiquitous.
The methodology of integrated nested Laplace approximations (INLA) provides a framework for performing Bayesian inference applicable to a large subclass of additive Bayesian hierarchical models. 
In combination with the stochastic partial differential equations (SPDE) approach it gives rise to an efficient method for spatial-temporal modeling.
In this work we build on the INLA-SPDE approach, by putting forward a performant distributed memory variant, INLA\textsubscript{DIST}, for large-scale applications. 
To perform the arising computational kernel operations, consisting of Cholesky factorizations, solving linear systems, and selected matrix inversions, we present two numerical solver options, a sparse CPU-based library and a novel blocked GPU-accelerated approach which we propose. 
We leverage the recurring nonzero block structure in the arising precision (inverse covariance) matrices, which allows us to employ dense subroutines within a sparse setting.
Both versions of INLA\textsubscript{DIST} are highly scalable, capable of performing inference on models with millions of latent parameters.
We demonstrate their accuracy and performance on synthetic as well as real-world climate dataset applications.
\end{abstract}


\section{Introduction}
\label{sec:intro}
Bayesian approaches to spatial-temporal modeling have shown to be a versatile tool to capture complex phenomena and handle imbalanced or missing observations~\cite{krainski2018advanced, joshuaetal2022bayesian}. 
Among their wide range of applications are climate and weather modeling~\cite{ arisido2017bayesian}, disease mapping~\cite{wah2020systematic, zhang2022application}, medical image analysis~\cite{lehnert2019large, mejia2020bayesian},  traffic management~\cite{batomen2020vulnerable, duan2021applying}, environmental changes~\cite{forlani2020joint, patterson2020probabilistic, senf2017bayesian} and health economic evaluations~\cite{heath2016estimating}.
We see growing amounts of data available to describe these phenomena and increasingly sophisticated models to characterize them~\cite{Allardetal2022nonsep, Salvanaetal2022advenctions}. 
This, however, poses computational challenges that often arise from the increase in dimensionality of the model~\cite{jurekKatzfuss2021multi, Nychkaetal2015multi}.

In this paper we present a GPU-accelerated distributed memory algorithm for performing approximate Bayesian inference on spatial-temporal models.
Our work is based on the methodology of integrated nested Laplace approximations (INLA)~\cite{rue2009approximate} which provides a framework for performing Bayesian inference. It is applicable to latent Gaussian models (LGMs), a large subclass of additive Bayesian hierarchical models. 
The INLA methodology leverages an analytic approximation scheme that makes use of nested second order approximations and sparse Gaussian Markov random field (GMRF) structures in the latent parameter space. 
This allows INLA to mitigate the prevailing problem of high-dimensional integration and enables efficient computations, while maintaining a high level of accuracy \cite{van2023new}.
The formulation of spatial and spatial-temporal models employs stochastic partial differential equations (SPDEs), inducing favorable theoretical and computational properties ~\cite{lindgren2011explicit}.
INLA is especially popular within the applied statistics communities as a user-friendly implementation exists in the form of an R package referred to as R-INLA~\cite{inlapkg}. 
The package offers a wide variety of modeling choices which allow for applications in many fields, see e.g.~\cite{gentrification2022, bhatt2015effect, de2019species, forlani2020joint,regExcessMort2022, konstantinoudis2021long, lu2018hierarchical, senf2017bayesian}. 
Its implementation is targeted towards single-node architectures, making use of many computational strategies for enhancing performance, ranging from algorithm design to a nested multi-threaded parallelism~\cite{gaedke2022parallelized}. 
While this is sufficient for many applications, there exists an abundance of statistical inference problems involving spatial-temporal data that remain very time-consuming if not computationally infeasible, requiring larger compute resources~\cite{Abdulahetal2022exageostatR}.

Spatial-temporal modeling is particularly challenging as the associated latent parameter space is very high-dimensional. 
The direct maximization of the likelihood using dense covariance matrices gives rise to large dense systems of equations. 
Handling them, in turn, has a large associated computational cost, which scales cubically with the number of observations~\cite{exageostat}.
Several methodological approaches have been proposed to avoid this computational burden. 
The simplest one is to consider separable space-time covariance functions, which result in space-time covariance matrices that are assembled as Kronecker products of purely temporal and spatial matrices. 
This restriction limits the choice to simplistic models but 
allows to
dramatically reduce the associated computational cost. 
There are other approaches
which also avoid constructing the full dense covariance
matrices~\cite{	Cressieetal2010stfixedrank, dataetal2016nearestneigh,Furreretal2006tapering,hamelijnck2021spatio, 
	katzfussG2021vecchia}. Alternative strategies consider
e.g.\ multi-resolution approaches~\cite{Katzfuss2017multi, Nychkaetal2015multi}, kernel
convolutions~\cite{rodriguesDiggle2010kernel} or low-rank
structure compression of the
Jacobian~\cite{10.1145/3428447, doi:10.1137/21M1466499}. 
A different perspective is taken by state-space approaches which
avoid working with dense covariance
matrices altogether, by considering the temporal
dynamics~\cite{CressieWikle2011book}.
They are, however, restricted in the class of available models as 
they do not accommodate for global variables. 
The SPDE  approach~\cite{lindgren2022spde, lindgren2011explicit,sigristKS2015spde}, on the other hand derives the spatial-temporal
precision matrix (inverse covariance matrix) from an SPDE
which is discretized using the Finite Element method. The
resulting precision matrices are naturally sparse (in contrast
to their dense inverses) while still allowing for fixed effects.
We consider a non-separable spatial-temporal
model, derived from an advection-diffusion
equation~\cite{lindgren2022diffusion}. 
This family of  physically inspired models captures,
more realistically, the spatial-temporal behavior of the
underlying physical process. 
We tackle the challenge of increased model complexity by developing a solution method which exploits the underlying model sparsity, utilizes parallelism and leverages the strengths of modern compute architectures.
Our novel algorithmic design of the INLA methodology for Bayesian spatial-temporal modeling contains the following main contributions: \\
\begin{itemize}
	\item The spatial-temporal SPDE formulation of the model induces a partially block-tridiagonal sparsity pattern in all of the arising high-dimensional precision matrices. We develop tailored solution methods that make use of this information.
	\item Bayesian inference problems that are formulated using precision instead of covariance matrices often require selected matrix inversions to extract marginal variances. 
	We derive a selected inversion scheme for block tridiagonal arrowhead (BTA) matrices, accommodating for the covariates related to the fixed effects of the model.
	\item We present a distributed-memory implementation of our method INLA\textsubscript{DIST}. It is part of a three-layer parallelism scheme that operates within and across CPU/GPU compute nodes.
	\item We integrate two alternative numerical solvers in  INLA\textsubscript{DIST} to handle the computational key operations, the sparse linear algebra library PARDISO~\cite{bollhofer2020state} and a novel BTA solver which we propose.	\item The BTA solver leverages GPU accelerators for the computational kernel operations. 
	\item We conduct a detailed performance analysis, demonstrating the improved performance and scaling behavior of our approach. In particular, we show that our approach scales linearly in time and independently of the number of observations. 
	\item We apply our method to a climate dataset~\cite{menne2012global} 
	over the US main continental area throughout the course of 1 year using more than $1$ million latent parameters and $2.5$ million observations.
\end{itemize}

\smallskip
The remaining paper is organized as follows. In Section~\ref{sec:background} we give an overview of the statistical concepts involved. Here, we first introduce Latent Gaussian models, INLA's applicable model class. 
Then we present the INLA methodology and the SPDE approach that is used for spatial-temporal modeling. 
Subsequently we introduce our main contributions. 
We start with an analysis of the algorithmic kernel operations and develop suitable solution methods in Section~\ref{sec:overall_algorithms}, followed by a discussion on their implementation which include multi-layer parallelism and GPU-based solvers in Section~\ref{sec:softwareAndImplementation}.
The next section is dedicated to numerical experiments on both synthetic and real-world datasets, demonstrating that we can solve large-scale Bayesian inference problems within tens of minutes. We conclude this work in Section~\ref{conclusion} with a summary.

\section{Background}
\label{sec:background}
\subsection{Latent Gaussian Models}
\label{sec:bay_hierarchical_models}
Bayesian hierarchical models are a popular multi-level approach in statistics to capture complex  phenomena due to their flexibility and interpretability ~\cite{banerjee2014hierarchical, congdon2014applied}.
We consider a large subclass called latent Gaussian models (LGMs) whose latent parameter space, as their name suggests, follows a multivariate normal distribution.
LGMs comprise a broad variety of commonly used statistical models, including regression models, penalized splines or random walk models, dynamic models like autoregressive models and spatial or spatial-temporal models~\cite{rue2009approximate}. 
The latter are part of the class of linear mixed-effect models. Their three-stage structure consists of a likelihood, a prior distribution of the latent parameter space and the hyperparameter space, respectively.
The observations $\vec{y}^T = (y_1,\, ...,\, y_m)^T$, are assumed to be conditionally independent given the parameters, that is
\begin{align}
	\vec{y} \ | \ \vec{\vec{\eta}}, \th_1 \sim \prod_{i=1}^{m} p (y_i | \eta_i, \th_1), 
	\label{eq:bhmlikelihood}
\end{align}
where $\th_1$ denote the hyperparameters associated to the likelihood and $\vec{\eta}$ the linear predictor which we define as $\vec{\eta} = \vec{Z} \vec{\beta} + \vec{A}\vec{u}$. 
The matrix $\vec{Z}$ contains the covariates and $\vec{\beta}$ the vector of fixed effects. The vector $\vec{u}$ represents the random effects with projection matrix $\vec{A}$. 
The observations are connected to the linear predictor through a link function $g$, where $\mathbb{E}(y_i \vert \eta_i, \th_1) = g(\eta_i) $.
The unknown vector-valued variables $\vec{\beta}$ and  $\vec{u}$ constitute the latent parameters $\vec{x} = (\vec{u}^T, \vec{\beta}^T)^T$.
It is assumed that $\vec{x}$ forms a Gaussian Markov random field (GMRF), whose precision matrix, i.e., its inverse covariance matrix, depends on the vector of hyperparameters $\th_2$. 
A GMRF with precision matrix $\vec{Q}$ and associated graph $\mathcal{G} = (\mathcal{V}, \mathcal{E})$ is a Gaussian field for which $\vec{Q}_{ij} \neq 0 \iff \{i,j\} \in \mathcal{E} \text{ for all } i \neq j$~\cite{rue2005gaussian}. 
The edges $\mathcal{E}$ encode the conditional dependence structure of the parameters. 
The graph $\mathcal{G}$ is assumed to be sparse and hence, the corresponding precision  matrix $\Q$ is as well.
Given a zero mean, the latent parameters $\x$ are distributed
\begin{equation}
	\vec{x} \, | \, \th_2  \sim \mathcal{N}(\vec{0}, \vec{Q}_{\x}(\th_2)^{-1}). 
	\label{eq:bhmlatent}
\end{equation}
We can combine the hyperparameters as  $\th = (\th_1, \th_2)$ and assign them to follow some prior distribution $	\vec{\theta} \sim p(\vec{\theta})$.
Thus, completing the three-stage hierarchical model 
for the observations $\y$, the latent parameters $\x$ and the hyperparameters $\th$.

\subsection{Fundamentals of the INLA methodology}
\label{sec:INLA_methodology}

The INLA methodology leverages an analytic approximation scheme to provide estimates of the posterior marginal distributions of the hyperparameters $p(\theta_i | \y)$ for all $i$ and of the latent parameters $p(x_j | \y)$ for all $j$. 
Other relevant statistics, such as credibility intervals and quantiles, can subsequently be derived.
It holds that
\begin{align}
	p(\theta_i | \vec{y} ) &= \iint p(\vec{x}, \th | \vec{y}) \, d\vec{x}\, d\th_{-i} = \int p (\th | y ) \, d \th_{-i}, \qquad \ \qquad \text{for all} \ i,
	\label{eq:post_hyperp} \\
	p(x_j | \vec{y} ) &= \iint p(\vec{x}, \th | \vec{y})\, d\vec{x}_{-j} \, d\th   = \int p(x_j | \th , \vec{y} ) p( \th | \vec{y} )\, d \th, \quad \ \text{for all} \ j.
	\label{eq:post_latent}
\end{align}
where $\th_{-i}$ denotes all hyperparameters except for the $i$-th one, and respectively for $\vec{x}_{-j}$. 
The integral over $\vec{x}$ is often very high-dimensional, due to the potentially large number of latent parameters $\x$, while the integral over the underlying hyperparameters $\th$ is assumed to be of much lower dimension~\cite{rue2017bayesian}. 
INLA approximates the joint posterior of the hyperparameters as

\begin{equation}
	p(\th|\vec{y}) = \frac{p(\vec{x},\th|\vec{y})}{p(\vec{x}|\th, \vec{y})} \propto \frac{p(\th)p(\vec{x}|\th)p(\vec{y}|\vec{x},\th)}{p(\vec{x}|\th, \vec{y})} \approx \frac{p(\th)p(\vec{x}|\th)p(\vec{y}|\vec{x},\th)}{p_G(\vec{x}|\th, \vec{y})} \bigg|_{\vec{x} = \vec{x}^*(\th)} \! := \tilde{p}(\th | \y).
	\label{eq:post_theta_approx}
\end{equation}
The equations can be derived from left to right as follows. The first identity is obtained through the chain rule. 
The proportionality holds by Bayes' rule, admitting the normalizing constant $p(\vec{y})$ which is independent of $\th$. 
The distribution $p_G$ represents a Gaussian approximation of $p(\vec{x}|\th, \vec{y})$, centered at the mode $\x^*(\th)$ of the full conditional  $p(\vec{x}|\th, \vec{y})$ for a fixed vector $\th$. More specifically, we assume that 
\begin{equation}
	\begin{aligned}
		p_G(\x | \th, \y) &= (2 \pi)^{-n/2} \vert \Q_{\x \vert \y} (\th) \vert ^{1/2}  \, \text{exp}(-\frac{1}{2}(\x^*(\th) - \x)^T \Q_{\x \vert \y}(\x^*(\th) - \x)), \ \text{where} \\
		\Q_{\x \vert \y}(\th) &:= \Q_{\x}(\th_2) + \vec{\tilde{A}}^T \vec{D} \vec{\tilde{A}}, 
		\quad \text{ with }  \vec{\tilde{A}} = [\vec{A}, \vec{Z}].
	\end{aligned}
	\label{eq:cond_latent_param}
\end{equation}
Here $\vec{D}$ denotes a diagonal matrix which is derived from a second-order Taylor expansion of the negative log-likelihood evaluated at the mode $\x^*(\th)$. If the the likelihood is normally distributed $\x^*(\th)$ can be determined directly by solving a linear system involving $\Q_{\x \vert \y}(\th)$, otherwise it is found iteratively~\cite{van2023new}. 
The mode $\th^*$ of $\tilde{p}(\th | \y)$ is not known a-priori, and poses an optimization problem.
It can be found using a quasi-Newton method maximizing over $\th$ which in each iteration requires the evaluation of  Eq.~(\ref{eq:post_hyperp}) and Eq.~(\ref{eq:cond_latent_param}) using the current value of $\th$.

Once $\th^*$ is determined, an exploration of $\tilde{p}(\th|\y)$ around the mode is performed  to set up an approximation scheme, using the evaluation points $\{\th^k\}^K_{k=1}$, from which the marginal distributions $p(\theta_i |\vec{y})$ for all $i$, are computed.
The posterior marginals of the latent parameters $p(x_j | \vec{y})$ for all $j$, are determined using Eq.~(\ref{eq:post_latent}), where now the only missing components are the $p(x_j | \th, \vec{y})$ for all $j$. 
There are a number of possibilities how to approximate them, each one presenting a different trade-off between accuracy and computational cost. 
The fastest but crudest strategy, known as empirical Bayes, uses only one integration point $\th^k$, namely at the mode $\th^*$. 
For more involved models this can, however, fail to capture skewness sufficiently or can generate bias.
There are two ways to further improve the accuracy. Firstly, by using additional integration points. Then the Gaussian approximation $p_G(\vec{x}|\th^k, \vec{y})$ is formed at each of them.
Additionally, it is possible to improve this approximation at every $\th^k$ using variational inference.
Here, the mean of each $p_G(\vec{x}|\th^k, \vec{y})$ is updated by adding a correction term that is determined through solving a variational problem~\cite{van2021correcting}.
The posterior marginal distributions $p(x_j | \vec{y} )$ are finally computed using information from each evaluation point $\{\th^k\}^K_{k=1}$ and the respectively chosen approximations of $p(x_j | \th, \vec{y} )$.  

\subsection{The Stochastic Partial Differential Equation Approach}


\label{sec:SPDE_approach}
To realistically represent spatial correlation in geostatistics one often uses Matérn covariance functions~\cite{stein2005spacetime}.
Let $u(s)$ be a Gaussian random field over a domain $D$ with $s \in D$. 
Then, as noted by Whittle \cite{whittle1954stationary}, the stochastic weak solution $u(s)$ to 
\begin{equation}
	\gamma_e(\gamma_s^2 - \Delta)^{\frac{\alpha}{2}} \  u(s)  = \mathcal{W}(s), \quad s \in \mathbf{R}^d
	\label{spde_matern}
\end{equation}
has a Matérn covariance function, assuming $\gamma_s, \alpha >0 $ are constants, $\Delta$ is the Laplace operator 
and $\mathcal{W}$ represents Gaussian white noise. 
Lindgren et al. show in~\cite{lindgren2011explicit} that Eq.~(\ref{spde_matern}) can be solved for $u(s)$ to find the spatial random effects of the linear mixed-effect model described in Sect.~\ref{sec:bay_hierarchical_models}. 
They directly relate the differential operator $L = (\gamma_s^2 - \Delta)^{\frac{\alpha}{2}}$ to the precision matrix of the spatial random effects.
The finite element method is employed to discretize Eq.~(\ref{spde_matern}).
This so-called SPDE approach has shown to be very attractive in spatial modeling, both from a theoretical as well as a practical point of view. It inherits consistent convergence properties and favorable computational aspects, such as sparsity in the precision matrix, from the finite element method.

For spatial-temporal models, we have a random field $u(s,t)$ containing the additional time component $t$, hence its covariance function $C(u(s_1,t_1), u(s_2,t_2))$ also has a dependency on time $t$. 
There are numerous approaches for a temporal extension of the Matérn covariance function, see e.g. \cite{jones1997models, stein2005spacetime}.
In all cases, the additional dimension clearly adds complexity to the model and hence, often separable covariance functions are chosen, i.e. $C(u(s_1,t_1), u(s_2,t_2)) = C_s(s_1, s_2) C_t(t_1, t_2)$, where $C_s$ represents the spatial and $C_t$ the temporal correlation. 
This simplifies the precision matrix of the corresponding model. 
However, separable models have a number of disadvantages, like smoothing property restrictions~\cite{stein2005spacetime}. 
Additionally, if one considers defining a model through the direct definition of its dynamics,
a separable model does not give rise to physically realistic dynamics~\cite{lindgren2022diffusion}.
Therefore, we rely on the non-separable spatial-temporal model extension suggested by Lindgren et al.~\cite{lindgren2022diffusion}. 
The authors present a diffusion-based family of models as an extension of the Matérn fields that additionally contains a first order derivative over time and smoothness parameters $(\alpha_t, \alpha_s, \alpha_e)$ which determine the order of the differential operators and thus the smoothness of the solution. 
The spatial-temporal random effects of the model employed in this work are then represented as the solution to the following SPDE 
\begin{equation}
	\left( \gamma_t \frac{\partial}{\partial t} + (\gamma_s^2 - \Delta) \right) \ u(s,t) = \mathcal{E}_{Q, \gamma_e}(s,t),
	\label{eq:demf_spde}
\end{equation}
where we assume $\alpha_t=1$, $\alpha_s=2$ and $\alpha_e=1$, for the general form see Sect.~2~\cite{lindgren2022diffusion}.
The SPDE has the non-negative scale parameters $(\gamma_s, \gamma_t, \gamma_e)$, 
the time derivative $\frac{\partial}{\partial t}$, the Laplace operator in space $\Delta$ and $\mathcal{E}_Q(s,t)$ which describes Gaussian noise that is uncorrelated in time but with an exponential correlation in space.

The stochastic weak solution of this SPDE gives rise to a Gaussian field with diffusive behavior. 
Lindgren et al.~\cite{lindgren2022diffusion} argue that the resulting covariance function is  the most natural as diffusion processes are fundamental to modeling spatial-temporal phenomena.
When restricting Eq.~(\ref{eq:demf_spde}) to only its spatial component one obtains a Matérn covariance function as in Eq.~(\ref{spde_matern}). 
The differential operator of Eq.~(\ref{eq:demf_spde}) directly relates to the precision operator of the random effects of the spatial-temporal model components and the parameters $(\gamma_s, \gamma_t, \gamma_e)$  are contained in the model's hyperparameters $\th$.
The discretisation of Eq.~(\ref{eq:demf_spde}) uses piece-wise linear basis functions in space and time. Thus, it gives rise to a continuous space-time solution while the underlying computations rely on sparse linear algebra operations. 

\section{Algorithmic Kernel Operations}
\label{sec:overall_algorithms}

\subsection{Overview}
\label{sec:methods}

After introducing the main statistical concepts and ideas behind the INLA-SPDE approach in the previous section, we now provide a more algorithmic overview that 
discusses the arising computational key components of our method and indicate the different opportunities for parallelism. 
We denote by $\tilde{p}(\th | \vec{y})$ the approximation to $p(\th | \vec{y})$ and respectively for other distributions.
The main steps are listed below. 

\vspace{0.3cm}
\begin{enumerate}
	\setcounter{enumi}{-1}
	\setlength\itemsep{0.25em}
	\item Construct scheme to evaluate $\tilde{p}(\th|\y)$ for fixed $\th$.
	\label{enum:INLA_step0}
	\item Solve optimization problem Eq.~(\ref{eq:post_theta_approx}) to determine the mode $\th^*$ of $\tilde{p}(\th|\y)$. \label{enum:INLA_step1}
	\item Compute the negative Hessian at the mode $\th^*$.  \label{enum:INLA_step2}
	\item Locate evaluation points in the neighborhood of $\th^*$ using the Hessian. Use numerical integration free algorithm to approximate $\tilde{p}(\theta_i|\vec{y})$. \label{enum:INLA_step3}
	\item Approximate densities $\tilde{p}(x_j | \th, \vec{y})$ for each evaluation point. \label{enum:INLA_step4}
	\item Combine information from each evaluation point to obtain $\tilde{p}(x_j |\vec{y})$ similarly to Step \ref{enum:INLA_step3}. \label{enum:INLA_step5}
\end{enumerate}
\vspace{0.3cm}
In the following we will discuss in more detail what each of the steps entails.
For simplicity we will assume the likelihood to follow a Gaussian distribution as the required kernel operations remain the same. All computations are performed in log-scale. To ease the notation we will define the function $f(\th)$ as
\begin{equation}
	f(\th) := - \, \text{log } \tilde{p}(\th | \y),
	\label{eq:f_def}
\end{equation}
where we consider the observations $\y$ as fixed in the current model.
\smallskip

\textbf{Step \ref{enum:INLA_step0}:} Evaluating $f(\th)$, and thus $\tilde{p}(\th | \y)$, efficiently for a fixed vector $\th$ is an integral part of the overall method and ubiquitous in the subsequent steps.
The evaluation of $f$ is split into its individual subcomponents, see Eq.~(\ref{eq:post_theta_approx}), that become additive through the introduced log-scale. 
While the likelihood and the prior of the hyperparameters are usually computationally cheap to evaluate, the prior $p(\x | \th)$ as well as the conditional distribution $p_G(\x | \th, \y)$ of the latent parameters are much more costly to compute. 
They give rise to three kernel operations which depend on $\th$ and particularly stand out: The computation of the log determinant of the precision matrix of the latent effects $\vec{Q}_{\vec{x}}(\th)$, the log determinant of the precision matrix from the conditional distribution in the denominator, $\vec{Q}_{\vec{x}| \vec{y}}(\th)$, as well as the computation of its conditional mean, which requires solving a linear system involving $\vec{Q}_{\vec{x}|\vec{y}}(\th)$. As these precision matrices are symmetric positive definite, it is most efficient to perform a Cholesky decomposition, use the diagonal entries of the factors to compute the log determinant and then, if required, perform a forward--backward substitution, to solve the linear system. 
In the spatial-temporal case, the dimension of the latent parameter vector $\x$, and thus the dimension of $\Q_{\x \vert \y}$, are directly related to the spatial-temporal discretization of the problem which grow quickly with increasing number of time steps or a finer resolution of the spatial domain.

\textbf{Step \ref{enum:INLA_step1}:} 
To find the minimum of $f$, or equivalently the maximum of Eq.~(\ref{eq:post_theta_approx}), a BFGS algorithm~\cite{nocedal2006numerical, BFGSsolver} is employed. As any quasi-Newton method it requires gradient information in every iteration to determine the next search direction. We estimate the gradient $\nabla f$ using a finite difference approximation as the analytical solution is not easily computable. 
Thus, every iteration of the optimization scheme does not only require a single function evaluation of $f(\th^l)$ for the current iterate $\th^l$ but instead necessitates the evaluation of of many $\th^l \pm \vec{\epsilon}_i$ that arise from the finite difference discretization to approximate the $i$-th directional derivative. The exact number of required function evaluations depends on the dimension of $\th$, $d(\th)$, as well as the chosen finite difference scheme.
While the optimization algorithm is by nature iterative and thus, sequential, the function evaluations required in each iteration for the gradient approximation can be completely parallelized as they are independent from each other. 
\smallskip

\textbf{Step \ref{enum:INLA_step2}:} The negative Hessian at the mode $\th^*$ is approximated using a second order finite difference scheme. This requires further evaluations of Eq.~(\ref{eq:post_hyperp}) which can be computed in parallel if resources allow.
\smallskip

\textbf{Step \ref{enum:INLA_step3}:}  
Approximation of the marginal posteriors of the hyperparameters $\tilde{p}(\theta_i | \y)$.
The space around of the mode $\th^*$ is explored according to the chosen integration strategy.
The inverse of the negative Hessian at the mode corresponds to a Gaussian approximation of the covariance matrix of $\th$. 
To correct for deviations from this Gaussian approximation, information from additional evaluation points can be employed, in a numerical integration free algorithm following~\cite{martins2013bayesian}, to compute the marginal posterior distributions $\tilde{p}(\theta_i | \y)$.
\smallskip

\textbf{Step \ref{enum:INLA_step4}:} 
Approximation of the conditional distributions $p(x_j| \th, \y)$. When the empirical Bayes integration strategy is used, the marginal means $\mu_j$ are directly deduced from the 
Gaussian approximation $p_G(\vec{x} | \th^*, \vec{y})$ at the mode. 
The marginal variances are the diagonal entries of the inverse of the precision matrix $\vec{Q}_{\x|\y}(\th^*)$ of $p_G$, i.e. $\Sigma_{jj} = (\vec{Q}^{-1}_{\x|\y}(\th^*))_{jj}$ for all $j$.
The inverse of a sparse matrix is generally dense, and therefore inversions of large sparse matrices are computationally expensive, if not infeasible, operations. In the case at hand we, however, only require the diagonal elements of the inverse.  There are specialized strategies to perform a partial or selected inversion that do not require the computation of all elements~\cite{JACQUELIN201884, doi:10.1137/09077432X}. A similar efficient and versatile selected inversion routine for sparse matrices which is based on Cholesky decomposition is known as Takahashi inversion~\cite{takahashi1973formation}. 
In Sect.~\ref{sec:blockInversion} we will put forward a different approach that is tailored to the particular sparsity pattern of the spatial-temporal precision matrices at hand to further improve performance. 
When an integration strategy is used that utilizes multiple evaluation points $\{\th^k\}_{k=1}^K$, the selected inversions are performed for each $\th^k$, and then used to construct $p(x_j| \th, \y)$. 
\smallskip

\textbf{Step \ref{enum:INLA_step5}:} 
The marginal posterior distributions $p(x_j | \y)$ are computed using the previously determined subcomponents $p(\th^k | \y)$ and $p(x_j | \th^k, \y)$ for all $k$.

\subsection{Sparsity Pattern of the Spatial-Temporal Precision Matrices}
\label{sec:sparsity_pattern_mat}

This section focuses on the recurring sparsity patterns that arise in the precision matrices $\Q_{\x}(\th)$ as part of $p(\x | \th)$ and $\vec{Q}_{\x | \y}(\th)$ as part of $p_G(\x | \th, \y)$ which is induced by the finite element discretization of Eq.~(\ref{eq:demf_spde}). 
For brevity we will implicitly assume the dependence of $\vec{Q}_{\x },  \vec{Q}_{\x | \y}$ on $\th$ for  from now on and omit it in the notation. 
We order the latent parameters $\x$ to first contain the random effects $\vec{u}$ and then the fixed effects $\vec{\beta}$, where $\vec{u}$ is internally structured to first accommodate all spatial grid points associated with the first time step, secondly all spatial grid points associated with the second time step, etc. This gives rise to a tridiagonal block structure in the precision matrices that relates to the discretized SPDE, where the off-diagonal blocks represent the coupling between two subsequent time steps. 
Each square block refers to one instance of the discretized spatial domain. 
The prior precision matrix of the latent parameters $\vec{\beta}$ is included in the last $n_b \times n_b$ block of $\Q_{\vec{x}}$, see Fig.~\ref{fig:Qst_schematic_matrix}.
The precision matrix $\Q_{\x| \y}$ additionally contains covariate information that arises from conditioning on the data $\y$, as defined in Eq.~(\ref{eq:cond_latent_param}). The sparsity pattern of the prior is maintained in the spatial-temporal part, but not in rows and columns associated to the fixed effects $\vec{\beta}$, which generally become dense, see Fig.~\ref{fig:Qxy_schematic_matrix}. 
The individual entries of the arising precision matrices are dependent on $\th$ whereas the sparsity pattern remains constant for a given set of covariates and spatial-temporal discretization, that is throughout all steps described in Sect.~\ref{sec:methods}.
	
	\begin{figure}[t]
		\centering
		\subfloat[Sparsity structure of the precision matrices $\vec{Q}_{\x}$. Each diagonal and off-diagonal block has size $n_s \times n_s$, except for the last one which is of size $n_b \times n_b$ and contains the precision matrix of the prior related to the fixed effects. On the main diagonal are $n_t$ blocks of size $n_s \times n_s$. By adding the final block $n_b \times n_b$ block this gives rise to a matrix a size $n = (n_s \cdot n_t + n_b) \times( n_s \cdot n_t + n_b)$.]
		{\label{fig:Qst_schematic_matrix}
			\begin{tikzpicture}[scale=0.5]
				\foreach \position in {(-6,6), (-5,5), (-4,4), (-3,3), (-2,2), (-1,1), (0,0), (1,-1)}
				\draw[fill=myorange] \position rectangle +(1,-1);
				
				\foreach \position in {(-6,5), (-5,4), (-4,3), (-3,2), (-2,1), (-1,0), (0,-1)}
				\draw[fill=myyellow] \position rectangle +(1,-1);
				
				\foreach \position in {(-5,6), (-4,5), (-3,4), (-2,3), (-1,2), (0,1), (1,0)}
				\draw[fill=myyellow] \position rectangle +(1,-1);
				
				\node[draw=none, text=white] at (3.15,-2.65) {$\vec{D}_{n_t+1}$};
				
				\draw[fill=myorange] (2,-2) rectangle (2.2,-2.2);
				
				
				
				
				\draw[decorate,decoration={brace, mirror}, xshift=-3pt] (-6,6) -- (-6,5.15) node[midway, left, xshift = -0.2]{$n_s$};
				
				\draw[decorate,decoration={brace, mirror}, xshift=-3pt] (-6,4.95) -- (-6,4) node[midway, left, xshift = -0.2]{$n_s$};
				
				\draw[decorate,decoration={brace}, yshift=3pt] (-6,6) -- (-5,6) node[midway, above]{$n_s$};
				
				
				
				\draw[decorate,decoration={brace}, xshift=3pt] (2.3,6) -- (2.3,-1.9) node[midway, right, yshift=-1pt, xshift=1pt]{$n_t \cdot n_s$};
				\draw[decorate,decoration={brace}, xshift=3pt] (2.3,-2) -- (2.3,-2.2) node[midway, right, yshift=-1pt, xshift=1pt]{$\, n_b$};
				
				
			\end{tikzpicture}
		} 
		\hspace{25pt}
		\subfloat[Sparsity structure of the precision matrices $\Q_{\x | \y}$, which are of the same size as (a), but in addition to the prior precision matrix $\Q_{\x}$ also contain information related to the data. The sparsity structure of the spatial-temporal component is not affected by conditioning on data while sparsity structure related to the fixed effects generally becomes dense.]
		{\label{fig:Qxy_schematic_matrix}
			\begin{tikzpicture}[scale=0.5]
				\foreach \position in {(-6,6), (-5,5), (-4,4), (-3,3), (-2,2), (-1,1), (0,0), (1,-1)}
				\draw[fill=myorange] \position rectangle +(1,-1);
				
				\foreach \position in {(-6,5), (-5,4), (-4,3), (-3,2), (-2,1), (-1,0), (0,-1)}
				\draw[fill=myyellow] \position rectangle +(1,-1);
				
				\foreach \position in {(-5,6), (-4,5), (-3,4), (-2,3), (-1,2), (0,1), (1,0)}
				\draw[fill=myyellow] \position rectangle +(1,-1);

				\draw[fill=myyellow] (-6,-2) rectangle (2,-2.2);
				\draw[fill=myyellow] (2,6) rectangle (2.2,-2);
				\draw[fill=myorange] (2,-2) rectangle (2.2,-2.2);
				
				\foreach \i in {0,1,2,3,4,5,6,7}
				\draw (-5+\i,-2) -- (-5+\i,-2.2);
				
				\foreach \i in {0,1,2,3,4,5,6,7}
				\draw (2, 5-\i) -- (2.2,5-\i );
				
				\node[draw=none, font=\small] at (-5.5,5.5) {$\vec{D}_1$};
				\node[draw=none, font=\small] at (-5.5,4.5) {$\vec{E}_1$};
				\node[draw=none, font=\small] at (-4.5,5.5) {$\vec{E}^T_1$};
				\node[draw=none, font=\small] at (-4.5,4.5) {$\vec{D}_2$};
				\node[draw=none, font=\small] at (-4.5,3.5) {$\vec{E}_2$};
				\node[draw=none, font=\small] at (-3.5,4.5) {$\vec{E}^T_2$};
				\node[draw=none, font=\small] at (-3.5,3.5) {$\vec{D}_3$};
				
				\draw[dotted] (-2.7,2.7) - - (-2.22,2.22) ;
				\draw[dotted] (-3.7,2.7) - - (-3.22,2.22) ;
				\draw[dotted] (-2.7,3.7) - - (-2.22,3.22) ;
				\draw[dotted] (0.3,-0.3) - - (0.75,-0.75) ;
				\draw[dotted] (0.3,-0.3) - - (0.75,-0.75) ;
				\draw[dotted] (1.3,-0.3) - - (1.75,-0.75) ;
				\draw[dotted] (0.3,-1.3) - - (0.75,-1.75) ;
				\node[draw=none, font=\small] at (1.5,-1.5) {$\vec{D}_{n_t}$};
				\node[draw=none, font=\small] at (3.15,-2.65) {$\vec{D}_{n_t+1}$};
				
				\node[draw=none, font=\small] at (2.95,5.5) {$\vec{F}^T_1$};
				\node[draw=none, font=\small] at (2.95,4.5) {$\vec{F}^T_2$};
				\draw[dotted] (2.8,3.7) -- (2.8,3.2) ;
				
				\node[draw=none, font=\small] at (-5.5,-2.65) {$\vec{F}_1$};
				\node[draw=none, font=\small] at (-4.5,-2.65) {$\vec{F}_2$};

				\draw[decorate,decoration={brace, mirror}, xshift=-3pt] (-6,6) -- (-6,5.15) node[midway, left, xshift = -0.2]{$n_s$};
				
				\draw[decorate,decoration={brace, mirror}, xshift=-3pt] (-6,4.95) -- (-6,4) node[midway, left, xshift = -0.2]{$n_s$};
				
				\draw[decorate,decoration={brace}, yshift=3pt] (-6,6) -- (-5,6) node[midway, above]{$n_s$};
				
				\draw[decorate,decoration={brace}, yshift=3pt] (-6,-2) -- (-5,-2) node[midway, above]{$n_s$};
				
				\draw[decorate,decoration={brace, mirror}, xshift=-3pt] (2,6) -- (2,5) node[midway, left, xshift = -0.2]{$n_s$};
				
				\draw[decorate,decoration={brace}, xshift=3pt] (2.2,6) -- (2.2,-1.9) node[midway, right, yshift=-1pt, xshift=1pt]{$n_t \cdot n_s$};
				\draw[decorate,decoration={brace}, xshift=3pt] (2.2,-2) -- (2.2,-2.2) node[midway, right, xshift = 1 pt, yshift=-1pt]{$n_b$};
				
				\draw[decorate,decoration={brace}, yshift=3pt] (2,6) -- (2.2,6) node[midway, above, xshift = 1 pt]{$n_b$};

			\end{tikzpicture}
		} 
		\caption{}
		\label{fig:sparsity_patterns}
	\end{figure}

	\subsubsection{Block Factorization and Inversion}
	\label{sec:blockInversion}
	In this section we develop blocked solution strategies using the structural matrix information discussed in the previous section.
	To compute the log determinants of $\Q_{\x|\y}$ and  $\Q_{\vec{x}}$, respectively, we perform block-wise Cholesky decompositions. 
	Given $\Q_{\x|\y}$ we start by computing the dense Cholesky decomposition of the first diagonal block $\L_{D_1} \L_{D_1}^T = \vec{D}_1$. Using this newly computed subfactor $\L_{D_1}$, we perform a triangular solve to obtain the remaining non-zero part of the Cholesky factor for the first $n_s$ columns. Subsequently, we update the next diagonal block $\vec{D}_2$ and  $\vec{D}_{n_t+1}$, as they are coupled to $\L_{E_1}$ and $\L_{F_1}$. Then, we can reiterate and perform a dense Cholesky decomposition of the next diagonal block $\vec{D}_2$. Algorithm~\ref{algo::cholesky} shows the of the resulting block factorization and a schematic overview of the sparsity pattern of the final Cholesky factor $\L_{\x| \y}$ is given in Fig.~\ref{fig:block_cholesky}. 
	For the prior precision matrix $\Q_{\vec{x}}$ the algorithm simplifies to factorizing a block tridiagonal matrix as the dense rows and columns referred to as $\vec{F}_i$ are equal to zero, see Fig.~\ref{fig:Qst_schematic_matrix}. Therefore all terms in Algorithm~\ref{algo::cholesky} which relate to $\vec{F}_i$ equate to zero.
	The blocked approach bears the advantage that each step of the iterative matrix factorization only requires the submatrices associated with the current time-step $t$ as well as the subsequent time-step $t+1$ at any given time, besides the small final diagonal block $\vec{D}_{n_t+1}$. 
	
	\begin{figure}
		\centering
		\begin{minipage}{.48\textwidth}
			\centering
			\begin{algorithm}[H]
				\caption{Block Cholesky \newline Factorization \label{algo::cholesky}}
				\begin{center}
						\begin{algorithmic}[1]		
							\FOR{$i = 1,2 \ldots n_t -1$}
							\STATE $\vec{L}_{D_i} = \text{chol}(\vec{D}_i)$
							\STATE $\vec{L}_{E_i} = \vec{E}_i \cdot  \vec{L}_{D_i}^{-T} $
							\STATE $\vec{L}_{F_i} = \vec{F}_i \cdot  \vec{L}_{D_i}^{-T}$
							\STATE $\vec{D}_{i+1} = \vec{D}_{i+1} - \vec{L}_{E_i} \cdot \vec{L}_{E_i}^T $
							\STATE $\vec{L_{F_{i+1}}} = \vec{L_{F_{i+1}}} - \vec{L}_{F_i} \cdot \vec{L}_{E_i}^T $
							\STATE $\vec{D}_{{n_t}+1} = \vec{D}_{n_t+1} -  \vec{L}_{F_i} \cdot \vec{L}_{F_i}^T$                     
							\ENDFOR      
							\STATE $\vec{L}_{D_{n_t}} = \text{chol}(\vec{D}_{n_t})$
							\STATE $\vec{L}_{F_{n_t}} = \vec{F}_{n_t} \cdot  \vec{L}_{D_{n_t}}^{-T}$
							\STATE $\vec{D}_{n_t+1} = \vec{D}_{n_t+1} -  \vec{L}_{F_{n_t}} \cdot \vec{L}_{F_{n_t}}^T$                     
							\STATE $\vec{L}_{D_{n_t+1}} = \text{chol}(\vec{D}_{n_t+1})$
						\end{algorithmic}
				\end{center}         
			\end{algorithm}
		\end{minipage}%
		\hspace{12pt}
		\begin{minipage}{.48\textwidth}
			\centering
			\begin{tikzpicture}[scale=0.58]				
				\draw[fill=none, white] (-6.5,6.5) rectangle (-6,6);
				
				\draw (-6,6) -- (2.2,-2.2);
				\draw(-6,6) -- (-6,5);
				\draw (2,-2.2) -- (2.2,-2.2);
				
				\fill[myorange] (-6,6) to  (2.2,-2.2) to  (1.2,-2.2) to  (-6,5) to cycle;

				\foreach \position in {(-6,5), (-5,4), (-4,3), (-3,2), (-2,1), (-1,0), (0,-1)}
				\draw[fill=myyellow] \position rectangle +(1,-1);
				
				\node[draw=none] at (-4.9,5.6) {$\L_{D_1}$};
				\node[draw=none] at (-3.9,4.6) {$\L_{D_2}$};
				\draw[dotted] (-3,3.7) - - (-2.52,3.22) ;
				
				\node[draw=none] at (-5.5,4.5) {$\L_{E_1}$};
				\node[draw=none] at (-4.5,3.5) {$\L_{E_2}$};
				\draw[dotted] (-3.7,2.7) - - (-3.22,2.22) ;
				\node[draw=none] at (2.3,-1.5) {$\L_{D_{n_t}}$};
				
				\draw[fill=myyellow] (-6,-2) rectangle (2,-2.2);

				\node[draw=none] at (3.1,-2.6) {$\L_{D_{n_t+1}}$};
				
				\foreach \i in {0,1,2,3,4,5,6,7}
				\draw (-5+\i,-2) -- (-5+\i,-2.2);
				
				\node[draw=none] at (-5.5,-2.6) {$\L_{F_1}$};
				\node[draw=none] at (-4.3,-2.6) {$\L_{F_2}$};
				\draw[dotted] (-3.6,-2.6) -- (-3.1,-2.6) ;
				
			\end{tikzpicture}
		\end{minipage}
		\caption{\textbf{Left-Panel: } Algorithm for Block Cholesky factorization of block tridiagonal arrowhead matrices $\Q_{\x \vert \y}$. \textbf{Right-Panel: } Corresponding sparsity pattern and labeled subblocks of the resulting Cholesky factor $\vec{L}_{\x \vert \y} \vec{L}_{\x \vert \y}^T = \Q_{\x \vert \y}$ .}
		\label{fig:block_cholesky}
	\end{figure}
	
	To efficiently compute the diagonal inverse elements $(\Q_{\x |\y}^{-1})_{ii}$ we derive a recursive strategy making use of the already computed Cholesky decomposition $ \L_{\x|\y} \L_{\x|\y}^T = \Q_{\x |\y}$ and its particular nonzero structure. Our approach is similar to methods used in quantum transport simulations where solutions to non-equilibrium Green's functions also necessitate selected inversions, see e.g.~\cite{10.1007/978-3-642-40047-6_54, LI20089408, 10.1145/3295500.3357156} as well as for Kalman-Bucy filtering~\cite{asif2000inversion}. In both cases the authors derive strategies to efficiently compute the block diagonal elements of the inverse of block tridiagonal matrices. We extend this to block tridiagonal arrowhead matrices, starting the derivation from the following identities
	\begin{equation}
		\Q = \S^{-1} = \L \L^T \iff \S = (\L \L^T)^{-1} \iff \S \L = \L^{-T}.
		\label{eq:SLisLinvT}
	\end{equation}
	We assume $\Q$ to be a symmetric positive-definite block tridiagonal arrowhead matrix. Its inverse $\S$ is generally dense but inherits the properties of symmetry and positive-definiteness. We follow the block notation given in Fig.~\ref{fig:Qxy_schematic_matrix} and write Eq.~(\ref{eq:SLisLinvT}) using a  submatrix notation. The inverse of an upper triangular matrix remains upper triangular. The blocks $\L_{D_{i}}^{-T}$ are the inverses of the individual block $\L_{D_{i}}^{T}$. The $*$ denotes unknown nonzero entries. 

		\small
		\begin{equation}	
			\renewcommand*{\arraycolsep}{0.05pt}
			\begin{bmatrix} 
				\S_{11} & \S_{12}   &  \hdots & \S_{1 n} & \S_{1 n_t +1}\\ 
				& \S_{22} & & &  \\
				\vdots & & \ddots & & \vdots  \\
				& & & \S_{n_t n_t } &   \S_{n_t n_t+1} \\
				\S_{n_t +1 1}	& \cdots &   &  \S_{n_t +1 n_t}  &  \S_{n_t+1 n_t+1}
			\end{bmatrix}  
			\cdot
			\begin{bmatrix} 
				\L_{D_1} & \vec{0}  &  & \cdots & \vec{0} \\ 
				\L_{E_1}	& \L_{D_2} & & & \\
				& \ddots & \ddots &   & \vdots \\
				& & \L_{E_{n_t -1}} & \L_{D_{n_t}} & \vec{0}  \\
				\L_{F_1}& \cdots & \L_{F_{n_t -1}} &    \L_{F_{n_t}} & \L_{D_{n_t +1}}
			\end{bmatrix}  
			=
			\begin{bmatrix} \L^{-T}_{D_1} &  * & * &  *& * \\ 
				\vec{0} & \L^{-T}_{D_2} & * & * &  * \\
				\vdots &  & \ddots  &  * & *  \\
				&  &  & \L^{-T}_{D_{n_t}} & *     \\
				\vec{0} &  \cdots  &  &  \vec{0}& \L^{-T}_{D_{n_t+1}}
			\end{bmatrix}   
			\label{eq:matrixNotSLisLinvT}
		\end{equation}
	\normalsize
	Using Eq.~(\ref{eq:matrixNotSLisLinvT}) we can extract the following identities. 
	\begin{alignat}{3}
		\S_{n_t+1 n_t+1}   \L_{D_{n_t+1}} =& \ \L^{-T}_{D_{n_t+1}}
		&&\Rightarrow	
		\S_{n_t+1 n_t+1} =\L^{-T}_{D_{n_t+1}} \L^{-1}_{D_{n_t+1}} 
		\label{eq:Snp1np1}\\
		\S_{n_t+1 n_t}  \L_{D_{n_t}}  +  \S_{n_t+1 n_t+1}  \L_{F_{n_t}} =& \ \vec{0}  &&\Rightarrow	
		\S_{n_t+1 n_t}=  -   \S_{n_t+1 n_t+1}  \L_{F_{n_t}} \L^{-1}_{D_{n_t}} 
		\label{eq:Snp1n}\\
		\S_{n_t n_t} \L_{D_{n_t}} +  \S_{n_t n_t +1} \L_{F_{n_t}}  =& \ \L^{-T}_{D_{n_t}} &&\Rightarrow  
		\S_{n_t n_t} \!= \L^{-T}_{D_{n_t}}  \L^{-1}_{D_{n_t}} \! - \! \S_{n_t n_t+1} \L_{F_{n_t}} \L^{-1}_{D_{n_t}} 
		\label{eq:Snn}
	\end{alignat}
	While for all $1 \leq i \leq n_t-1$ we can derive the following.
	\begin{align}
		\S_{ii} \L_{D_{i}} +  \S_{i i+1} \L_{E_i} &+ \S_{i n_t+1} \L_{F_i}  = \L^{-T}_{D_i} \nonumber \\ &\Rightarrow  \S_{ii} = \L^{-T}_{D_i} \L^{-1}_{D_{i}} - \S_{i i+1} \L_{E_i}  \L^{-1}_{D_{i}}  - \S_{i n_t+1} \L_{F_i} \L^{-1}_{D_{i}} 
		\label{eq:Sii}\\
		\S_{i+1 i} \L_{D_{i}} + \S_{i+1 i+1} \L_{E_i} &+ \S_{i+1 n_t+1} \L_{F_i} = \vec{0} \nonumber \\ &\Rightarrow \S_{i+1 i} = - \S_{i+1 i+1} \L_{E_i} \L^{-1}_{D_{i}}  - \S_{i+1 n+1} \L_{F_i}   \L^{-1}_{D_{i}}  
		\label{eq:Sip1i}\\
		\S_{n_t+1 i} \L_{D_{i}} \! + \! \S_{n_t+1  i+1} \L_{E_i} &+  \S_{n_t +1 n_t +1}  \L_{F_i} = \vec{0} \nonumber \\ &\Rightarrow \! \S_{n_t+1 i} =\! - \S_{n_t+1i+1} \L_{E_i}\L^{-1}_{D_{i}} \!- \! \S_{n_t +1 n_t +1}  \L_{F_i} \L^{-1}_{D_{i}} 
		\label{eq:Snp1i}
	\end{align}
	Taking Eq.~(\ref{eq:Sii}) and substituting the unknown terms for Eq.~(\ref{eq:Sip1i}) and Eq.~(\ref{eq:Snp1i}), as well as using the fact that $\S_{ij}^T = \S_{ji}$, one obtains
	\begin{equation}
		\begin{aligned}
			\S_{ii} = \L^{-T}_{D_i} \L^{-1}_{D_{i}} - (- &\S_{i+1 i+1} \L_{E_i} \L^{-1}_{D_{i}}  - \S_{i+1 n_t +1} \L_{F_i}   \L^{-1}_{D_{i}} )^T \L_{E_i}  \L^{-1}_{D_{i}} \\
			-& (- \S_{n_t +1i+1} \L_{E_i}\L^{-1}_{D_{i}} -  \S_{n_t +1 n_t +1}  \L_{F_i} \L^{-1}_{D_{i}})^T \L_{F_i} \L^{-1}_{D_{i}},
		\end{aligned}
	\end{equation}
	and after rearranging 
	\begin{equation}
		\begin{aligned}
			\S_{ii} = \L^{-T}_{D_i}(\vec{I} +  \L^T_{E_i}\S_{i+1 i+1} \L_{E_i} + \L_{F_i}^T &\S_{n_t +1 n_t +1}  \L_{F_i} \\
			+  \  \L^T_{F_i} \S_{n_t +1i+1 } &\L_{E_i} +  \L_{E_i}^T \S_{i+1n_t +1}  \L_{F_i}) \L^{-1}_{D_{i}}.
		\end{aligned}
	\end{equation}
	Using the above equations we can deduce an efficient algorithm, see Alg.~\ref{alg:blockInv}, to recursively compute the blocks $\S_{ii}$ starting at $i \! = \! n_t + 1$, performing an upward traversal through the matrix.
	During each iteration $i$ we make use of the previously computed diagonal blocks $\S_{i+1i+1}$ and $\S_{n_t+1n_t+1}$. One additionally requires the off-diagonal blocks $\S_{n_t+1 i}$ for $2 \leq i \leq n_t$  that are computed recursively using Eq.~(\ref{eq:Snp1n}) and~(\ref{eq:Snp1i}). 
	
	\begin{algorithm}[H]
		\caption{Selected Block Inversion \label{algo::selected_inversion}}
		\begin{center}
				\begin{algorithmic}[1]		
					\STATE $\S_{n_t+1n_t+1} = \vec{L}_{D_{n_t+1}}^{-T} \cdot  \vec{L}_{D_{n_t+1}}^{-1}$
					\STATE $\S_{n_t+1 n_t} = -   \S_{n_t+1 n_t+1}  \L_{F_{n_t}} \L^{-1}_{D_{n_t}}$ 
					\STATE $\S_{n_t n_t} = \L^{-T}_{D_{n_t}}  (\vec{I}  +  \L^T_{F_{n_t}} \S_{n_t+1 n_t+1}  \L_{F_{n_t}}) \L^{-1}_{D_{n_t}}$
					\FOR{$i = n_t -1, n_t -2,  \ldots, 1$}
					\STATE 	$\S_{ii}\!=\!\L^{-T}_{D_i}(\vec{I}\!+\!\L^T_{E_i}\S_{i+1 i+1} \L_{E_i}\!+\!\L_{F_i}^T \S_{n_t +1 n_t +1}\L_{F_i}\!+  \!\L^T_{F_i} \S_{i+1 n_t +1} \L_{E_i} +  \L_{E_i}^T \S_{n_t+1 i+1}  \L_{F_i}) \L^{-1}_{D_{i}}$
					\STATE $\S_{n_t +1 i} = - (\S_{n_t +1 i+1} \L_{E_i} +  \S_{n_t+1 n_t +1}  \L_{F_i}) \L^{-1}_{D_{i}}$
					\ENDFOR      
				\end{algorithmic}
		\end{center}         
		\label{alg:blockInv}
	\end{algorithm}
	
	This blocked approach bears, again, the advantage that each step of the iterative selected block inversion only requires the submatrices associated with the current time-step $t$ as well as the previously computed time-step $t+1$ at any given time, besides the initially computed small diagonal block $\S_{n_t+1n_t+1}$. 
	
	\subsection{Sparse Factorization and Inversion}
	Alternatively to using dense block operations for the computational kernel tasks, it is also possible to employ entirely sparse solution methods.
	Each of the diagonal blocks of size $n_s \times n_s$ as well as the off-diagonal blocks are themselves sparse, since they are obtained from a finite element discretization. Their sparsity pattern is, however, not known a-priori as it depends on the particular spatial discretization and node ordering.
	While sparse Cholesky decomposition is a widely supported functionality that is part of many sparse linear algebra libraries, efficient selected inversion routines are not frequently supported. 
	One possibility is to employ the state-of-the-art sparse direct solver PARDISO~\cite{bollhofer2020state}.
	It performs efficient Cholesky factorizations through matrix permutations which allow for internal thread parallelization and lead to drastically reduced fill-in. 
	It additionally provides a selected matrix inversion routine building on the work of Takahashi~\cite{takahashi1973formation}. We make use of the PARDISO library as an entirely CPU-based alternative to our previously derived blocked approach.
	
	\subsection{Complexity}
	\label{sec:complexity}
	We dedicate this section to discussing the computational cost of  the proposed block Algorithms~\ref{algo::cholesky} and~\ref{alg:blockInv}, which we refer to as block tridiagonal arrowhead (BTA) approach. We compare it to the respective alternatives, entirely dense routines and entirely sparse approaches as employed in PARDISO.
	An overview for $\Q_{\vec{x}}, \Q_{\x \vert \y} \in \mathbb{R}^{n \times n}$ with $n = n_s \cdot n_t + n_b$, can be found in Table~\ref{tab:complexity}. 
	The dense version of both algorithms scale cubically in $n$. 
	The BTA approach takes the block tridiagonal structure of $\Q_{\vec{x}}$ into account, and therefore the cost for the factorization (Alg.~\ref{algo::cholesky}) or selected block inversion  (Alg.~\ref{alg:blockInv}) scales linearly in $n_t$ but cubically in $n_s$. 
	The matrix $\Q_{\vec{x} \vert \y}$ contains the additional arrowhead structure. Therefore there is an extra cost associated to those mixed terms, which is however, relatively small when $n_b < \! < ns$, as in our case.
	For PARDISO, we have that while using planar spatial meshes and by using nested dissection~\cite{doi:10.1137/0710032} the matrix $\Q_{\vec{x}}$ can be factorized with a complexity of $O( (n_s \cdot n_t)^{1.5}+n_b^{3})$. This nested dissection method can be also applied to reorder $\Q_{\x \vert \y}$ thus resulting  in a complexity of $O( (n_s \cdot n_t)^{1.5} + (n_s \cdot n_t+1)n_b^{3} )$. 
	
	\begingroup
	\renewcommand{\arraystretch}{1.3} 
	\begin{table}[t]
		\centering
		\begin{tabular}{l| c|c|c}
			\multirow{2}{*}{}
			& \textbf{DENSE} & \textbf{BTA} & \textbf{PARDISO}  \\ 
			& CHOL / INV. & CHOL / BLK. INV. & CHOL. / SEL. INV.  \\ \hline \hline 
			$\Q_{\x}$ & $O((n_s n_t + n_b)^3) $ & $O(n_s^3 n_t + n_b^3)$  & $O( (n_s n_t)^{3/2}+n_b^{3})$ \\ \hline 
			$\Q_{\x \vert \y }$ & $O((n_s  n_t + n_b)^3) $  & $O(n_s^3 n_t + (n_s n_t+1)n_b^{3})$  &  $O( (n_s n_t)^{3/2} + (n_s n_t+1)n_b^{3} )$ \\ \hline \hline
		\end{tabular}
		\caption{Overview of the computational complexity for the required Cholesky factorization and matrix inversion of  $\Q_{\vec{x}}$ \& $\Q_{\x \vert \y}$, see Fig.~\ref{fig:sparsity_patterns}. We compare a standard dense approach with our proposed BTA approach and the sparse solver PARDISO.
			Here, $n_s$ denotes the spatial block size, $n_t$ the number of time steps and $n_b$ the number of fixed effects.
	}
	\label{tab:complexity}
\end{table}
\endgroup

It is additionally worth noting that the dimension of $\Q_{\x \vert \y}$ is determined by the number of latent parameters in the model and does not change with the number of available observations. Similarly, the sparsity pattern of the spatial-temporal component remains unaffected with increasing numbers of observations. The remaining part of the matrix is generally assumed to be dense, and thus the complexity of all approaches, and in particular Algorithm~\ref{algo::cholesky} and~\ref{alg:blockInv} are independent of the number of observations. 

\section{Software and Implementation}
\label{sec:softwareAndImplementation}

We developed a C$++$\,--based implementation of our previously introduced INLA-SPDE approach for spatial-temporal modeling, which we will refer to as INLA\textsubscript{DIST}.
It makes use of the Eigen library~\cite{eigenweb} for the basic linear algebra operations and data structure management. We employ a BFGS template library provided by~\cite{BFGSsolver} to solve the optimization problem arising in Sect.~\ref{sec:methods}, Step~\ref{enum:INLA_step1}. 
Our implementation supports a combination of distributed and shared memory parallelism with the details presented in subsequent section. 
INLA\textsubscript{DIST} can be run with two different numerical solvers for the computational kernel operations. 
One option is the sparse solver PARDISO. The combination of INLA\textsubscript{DIST} and PARDISO will be referred to as INLA\textsubscript{PARDISO}. 
Secondly, we put forward a GPU-accelerated block tridiagonal arrowhead (BTA) solver implementation, based on Algorithm~\ref{algo::cholesky} and~\ref{alg:blockInv}, in combination with INLA\textsubscript{DIST} this will be referred to as INLA\textsubscript{BTA}. 

\subsection{Multi-layer Parallelism}
\label{sec:parallel_approach}
INLA\textsubscript{DIST} includes a multi-layer parallel scheme that ranges from algorithm-specific independently executable tasks to parallelizable linear algebra operations.
Starting from the coarsest level of parallelism, we can observe that in various stages of the algorithm, as presented in Sect.~\ref{sec:methods}, one requires evaluations of Eq.~(\ref{eq:f_def}) for different values of $\th$, see e.g. Step~\ref{enum:INLA_step1} and \ref{enum:INLA_step2}. 
The former consists of the optimization phase, where one requires not only an approximation to Eq.~(\ref{eq:f_def}) but also its gradient for every iteration of the quasi-Newton method. 
As we use a finite difference approximation, we can perform the necessary function evaluations in parallel to obtain an estimate of the gradient. A first order central difference approximation is used and hence the number of necessary evaluations is $2 \cdot d(\th) +1$, where $d(\th)$ describes the dimension of the hyperparameters $\th$. 
In Step~\ref{enum:INLA_step2}, numerous further parallel function evaluations are required for the second order finite difference approximation of the Hessian at the mode. Subsequently, and depending on the integration strategy, there are additional parallelizable function evaluations needed around the mode of Eq.~(\ref{eq:post_hyperp}), i.e. Step~\ref{enum:INLA_step3}. 
Likewise, the partial matrix inversions can be performed simultaneously in Step~\ref{enum:INLA_step4}. 
In all cases the function evaluations are embarrassingly parallel. 
We have implemented them using a distributed memory approach, where each task is executed by a separate process.

\begin{figure}
	\centering
	\includegraphics[width=\textwidth]{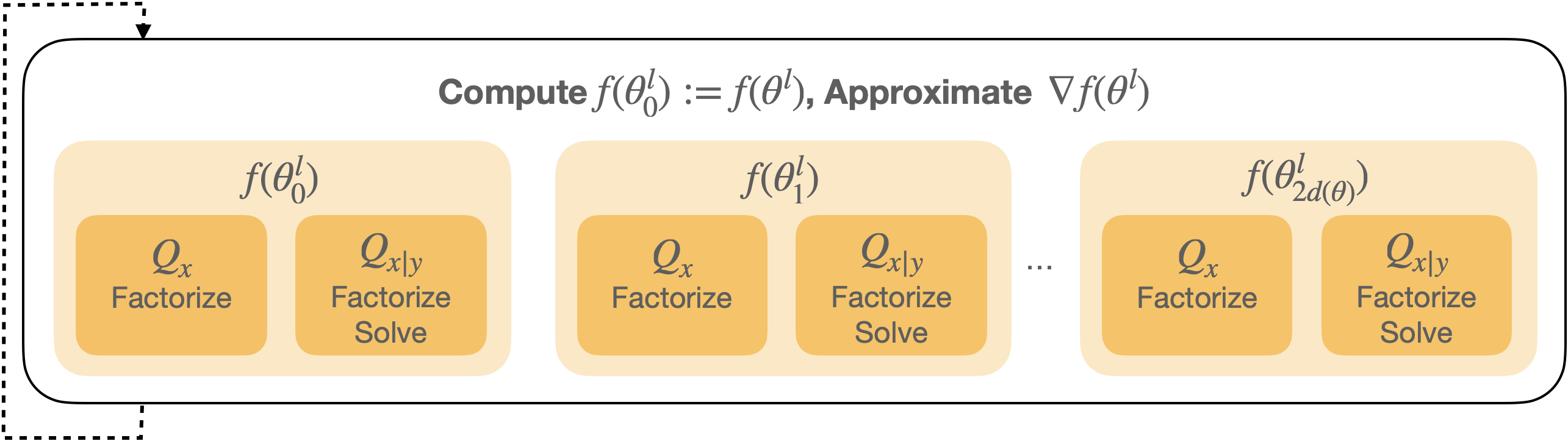}
	\caption{Schematic overview of the parallel scheme during one iteration of the optimization phase (for simplicity, without the line search routine). The finite difference gradient approximation entails $2 \cdot d(\th) + 1$ independent function evaluations of $f$ which are executed by separate processes. 
		Each process contains multiple threads that can execute the Cholesky factorizations of $\Q_{\x}$ and $\Q_{\x \vert \y}$ independently of each other. Alternatively each function evaluation is executed by 2 processes such that  $\Q_{\x}$ and $\Q_{\x \vert \y}$ can be factorized by separate processes.
		Each factorization itself is then parallelized by employing parallel linear algebra operations.}
	\label{fig:overview_parallel_scheme}
\end{figure}

Further possibilities for parallelism present themselves within every function evaluation of $f$ for a fixed parameter configuration of $\th$. Each term in the quotient can be evaluated independently, where especially the Cholesky factorizations of the precision matrix $\Q_{\vec{x}}$ belonging to $p(\x \vert \th)$ and of the precision matrix $\Q_{\x \vert \y}$ belonging $p_G(\x \vert \th, \y)$ stand out as computationally demanding, see Step~\ref{enum:INLA_step0}. The factorizations of the two different matrices are independent of each other, and thus parallelizable.
This is realized, either, through a shared-memory approach using multiple threads within each node or, again, as a distributed approach, and is chosen according to the compute node architecture. More specifically the arising tasks are separated into computing the numerator and denominator of Eq.~(\ref{eq:post_theta_approx}), respectively. 
Additionally, the linear algebra operations of each matrix factorization and selected inversion (as well as additional standard matrix operations) are parallelized through the employed numerical solvers and make use of shared memory approaches.
A schematic overview of the parallel scheme during the optimization phase can be found in Fig.~\ref{fig:overview_parallel_scheme}.

\subsection{BTA Solver Implementation}
\label{sec:BTA_implementation}
We developed a block tridiagonal arrowhead (BTA) solver that implements Algorithm~\ref{algo::cholesky} and~\ref{algo::selected_inversion}.
It makes use of accelerators to leverage the power and efficiency of GPUs in performing dense block operations.
The required kernel operations are performed by standard state-of-the-art linear algebra libraries, namely by MAGMA~\cite{al2020magma}, cuBLAS~\cite{cuda} and LAPACK~\cite{lapack}. 
Depending on the compute infrastructure and the problem size the complete Cholesky factor can exceed the GPU memory.
To nevertheless allow for almost arbitrarily large parameter spaces only the currently required submatrix blocks are iteratively copied to GPU memory to compute the Cholesky factorization.
The log determinants are then directly computed on GPU. 
When necessary, the complete Cholesky factor is stored in main memory, in which case the forward-backward substitution is then performed on CPU. 
To compute the selected block inversion the necessary submatrices of the Cholesky factor are recursively copied back to GPU memory throughout the computation as needed.

\section{Numerical Experiments}
\label{experiments}
To demonstrate the performance of our method we will present two numerical case studies, a family of synthetic datasets of varying sizes and a real-world application concerned with continuous air temperature modeling over the United States. 
We will show that our implementation of the INLA-SPDE approach INLA\textsubscript{DIST} is highly scalable, and can estimate model the parameter distributions of large-scale problems within a reasonable amount of time.
The numerical experiments using were carried out at the Erlangen National High Performance Computing Center (NHR@FAU).  The simulations using INLA\textsubscript{BTA} were conducted on GPGPU nodes, each equipped with 64-core dual chip AMD EPYC 7713 “Milan” processors @ 2.0 GHz and 8 Nvidia A100 GPUs. The simulations using INLA\textsubscript{PARDISO} were conducted on dual socket Intel Xeon Platinum 8360Y “Ice Lake” processors (36 cores per chip) @ 3.6 GHz.
If not stated otherwise each instance of PARDISO was called using 32 threads. This represents the most performant choice, as we will show in Sect.~\ref{Sec:KernelOps}.

\subsection{Synthetic Dataset}
\label{sec:synthetic_dataset}
We generate a family of synthetic datasets based on the model class described in Sect.~\ref{sec:bay_hierarchical_models}, that is
\begin{equation}
	\y = \vec{Z} \vec{\beta} + \vec{A} \vec{u} + \vec{\epsilon}, \ \text{ where } \ \epsilon_i \sim \mathcal{N}(0, \gamma_e^{-1}). 
	\label{eq:linPred_SynDat}
\end{equation}
In the following we outline how the individual model components of Eq.~(\ref{eq:linPred_SynDat}) were chosen.
The vector $\vec{\beta}$ contains the fixed effects of the model which include an intercept. 
We chose a random 6-dimensional vector, as the number of fixed effects is typically small, with values between $[-5,5]$.
We sample covariates $\vec{Z}$ stemming from linear as well as nonlinear functions with additional random uniform noise to interact with the fixed effects. 
The vector $\vec{u}$ is associated to the spatial-temporal field. Exemplary we consider the entire globe as our spatial domain over equidistant time units. The discretization is done using first order finite elements as previously described. 
The random field $\vec{u}$ is sampled from the spatial-temporal component of  $\Q_{\x}(\th)$ with mean zero, as by model definition. 
The precision matrix $\Q_{\x}(\th)$ is formed by the discretization of Eq.~(\ref{eq:demf_spde}) and independent prior variances for each fixed effect. 
It is parametrized by the 4-dimensional hyperparameter vector $\th = (\gamma_e, \gamma_s, \gamma_t, \gamma_u)^T$, which includes a noise term for the observations and the scaling parameters of the SPDE. They hyperparameters are chosen such that they give rise to a realistic system.
We sample random locations over the globe from a uniform distribution to represent the measurement stations. 
They generally do not coincide with the nodes of the finite element mesh. For simplicity we assume the stations to remain constant over time, which is, however, not a requirement, as each measurement is projected onto the finite element mesh using the projection matrix $\vec{A}$. 
Twice as many observations per time step are sampled as there are spatial mesh nodes.

Additionally, we select a prior for the hyperparamters $\th$. We choose a penalized complexity prior \cite{simpson2017penalising} as they are particularly suited for additive models, allow for explicitly incorporating probability statements on the parameters, and penalize unnecessary complexity in the model, as its name suggests~\cite{krainski2018advanced}.
We generate datasets of varying mesh sizes in time and space, thus giving rise to differently sized problems, see Table~\ref{tab:syn_base_case}. 
For both, the Base Case (BC) model and Temporal Scaling Case III (TS III), we have latent parameter spaces with more than $1$ million unknown parameters and more than $2$ million observations.

\begin{table}[t!]
	\centering
	\begin{tabular}{ll|r|r|r|r|r}
		
		&	& $n_s$& $n_t$ & $n_b$& $n_o$ & $n$  \\ \hline \hline
		\textbf{Base Case} & BC  &\textbf{4002} &\textbf{250} & \textbf{6} & \textbf{2 001 000} & \textbf{1 000 506} \\ \hline \hline
		\multirow{3}{*}{Temporal  Scaling}
		& TS I & 4002 &  50 & 6 & 400 200 & 200 106   \\ \cline{3-7}
		& TS II & 4002 &  100 & 6 & 800 400 & 400 206 \\ \cline{3-7}
		& TS III & 4002 &  500 & 6 & 4 002 000 & 2 001 006 \\ \hline \hline
		\multirow{4}{*}{Spatial Scaling}
		& SS I &  4002 &  30 & 6 & 240 120 & 120 066    \\ \cline{3-7}
		& SS II & 10242 &  30 & 6 & 614 520 &  307 266 \\ \cline{3-7}
		&  SS III  & 16002 &  30 & 6 & 960 120 & 480 066 \\ \cline{3-7}
		& SS IV & 20252 &  30 & 6 & 1 215 120 & 607 566 \\ \hline \hline
		\hspace{1pt}
	\end{tabular}
	\centering
	\caption{Differently sized model problems and their corresponding key dimensions, where $n_s$ denotes the number of spatial nodes, $n_t$ the number of time steps, $n_b$ the number of fixed effects, $n_o$ the number of observations and $n$ the number of latent parameters, thus $\Q_{\x}, \Q_{\x \vert \y} \in \mathbb{R}^{n \times n}$ with $n = n_s \cdot n_t + n_b$.}
	\label{tab:syn_base_case}
\end{table}

\subsubsection{Convergence Analysis BFGS}

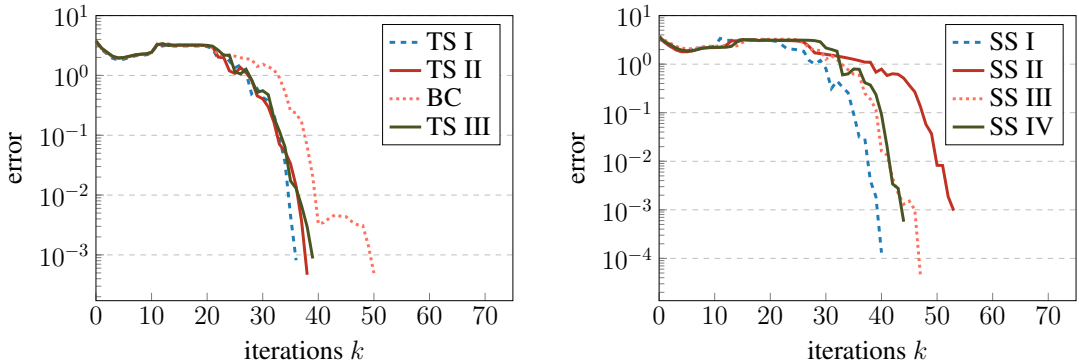
\begin{figure}[t!]
	\centering
	\begin{tikzpicture}[scale=0.7,font=\Large]
		\begin{axis}
			[legend pos={north east}, 
			xlabel = {iterations $k$}, 
			ymax = 10,
			ymode = log,
			ylabel = {error}, 
			xmin=0, xmax=75,
			ylabel style={yshift=0.3cm},
			xlabel style={yshift=-0.1cm},
			ymajorgrids=true, grid style=dashed, xtick pos=left, ytick
			pos=left, every axis plot/.append style={ultra thick},
			width=9.5cm,
			height=7cm,
			legend cell align={left},]
			
			\pgfplotstableread{benchmarks/bfgs_results_RGF_ns4002_nt50_nb6_9_2_1_fixedData_delta1e-5_past1_relEps1e-4.txt}\loadedtable
			\addplot [myblue, dashed]table[x="iter", y="error"] {\loadedtable};	
			
			\pgfplotstableread{benchmarks/bfgs_results_RGF_ns4002_nt100_nb6_9_2_1_fixedData_delta1e-5_past1_relEps1e-4.txt}\loadedtable
			\addplot [myred]table[x="iter", y="error"] {\loadedtable};				
			
			\pgfplotstableread{benchmarks/bfgs_results_RGF_ns4002_nt250_nb6_9_2_1_fixedData_delta1e-5_past1_relEps1e-4.txt}\loadedtable
			\addplot [bittersweet, dotted]table[x="iter", y="error"] {\loadedtable};	
			
			\pgfplotstableread{benchmarks/bfgs_results_RGF_ns4002_nt500_nb6_9_2_1_fixedData_delta1e-5_past1_relEps1e-4.txt}\loadedtable
			\addplot [armygreen]table[x="iter", y="error"] {\loadedtable};	
			
			\legend{TS I, TS II , BC, TS III},
			
		\end{axis}
	\end{tikzpicture}
	\hspace{15pt}
	\begin{tikzpicture}[scale=0.7,font=\Large]
		\begin{axis}
			[legend pos={north east}, 
			xlabel = {iterations $k$}, 
			ymax = 10,
			ymode = log,
			ylabel = {error}, 
			xmin=0, xmax=75,
			ylabel style={yshift=0.3cm},
			xlabel style={yshift=-0.1cm},
			ymajorgrids=true, grid style=dashed, xtick pos=left, ytick
			pos=left, every axis plot/.append style={ultra thick},
			width=9.5cm,
			height=7cm,
			legend cell align={left},]
			
			\pgfplotstableread{benchmarks/bfgs_results_RGF_ns4002_nt30_nb6_9_2_1_fixedData_delta1e-5_past1_relEps1e-4.txt}\loadedtable
			\addplot [myblue, dashed]table[x="iter", y="error"] {\loadedtable};	
			
			\pgfplotstableread{benchmarks/bfgs_results_RGF_ns10242_nt30_nb6_9_2_1_fixedData_delta1e-5_past1_relEps1e-4.txt}\loadedtable
			\addplot [myred]table[x="iter", y="error"] {\loadedtable};				
			
			\pgfplotstableread{benchmarks/bfgs_results_RGF_ns16002_nt30_nb6_9_2_1_fixedData_delta1e-5_past1_relEps1e-4.txt}\loadedtable
			\addplot [bittersweet, dotted]table[x="iter", y="error"] {\loadedtable};	
			
			\pgfplotstableread{benchmarks/bfgs_results_RGF_ns20252_nt30_nb6_9_2_1_fixedData_delta1e-5_past1_relEps1e-4.txt}\loadedtable	
			\addplot [armygreen]table[x="iter", y="error"] {\loadedtable};	
			
			\legend{SS I, SS II, SS III, SS IV},
			
		\end{axis}
	\end{tikzpicture}
	\caption{The $x$-axis shows the number of iterations over the error on the $y$-axis, defined as $\| \th^* - \th^k\| $, where $\th^*$ denotes the optimum, $\th^k$ refers to the $k$-th iterate, and $\| \cdot \|$ the Euclidean norm. The results for the different spatial-temporal model sizes from Table~\ref{tab:syn_base_case} are shown. The same initial values were chosen for all datasets. \textbf{Left-Panel:} Model problems related to the temporal scaling, including the Base Case.  \textbf{Right-Panel:} Model problems related to the spatial scaling.  
	}
	\label{fig:bfgs_convergence}
\end{figure}

To solve the optimization problem described in Sect.~\ref{sec:INLA_methodology}, Step~\ref{enum:INLA_step1}, a BFGS-algorithm in combination with a backtracking line search using Wolfe condition is employed. 
In Fig.~\ref{fig:bfgs_convergence} we present convergence results for different cases from Table~\ref{tab:syn_base_case}. It can be seen that the overall convergence behavior for each of the differently sized problems is similar while the exact number of iterations varies. In all cases the same initial guess was used. 

\subsubsection{Single Node Performance}
We are not aware of any distributed software library that supports the model types treated in this work. The 
R-INLA package\footnote{through the INLAspacetime package (https://github.com/eliaskrainski/INLAspacetime)}, however, provides a single node implementation which, since~\cite{gaedke2022parallelized}, supports nested multi-threading.
Therefore, we conduct a performance comparison between R-INLA and our CPU-based INLA\textsubscript{PARDISO} implementation, using a set of smaller models but following the same setup as described in Sect.~\ref{sec:synthetic_dataset},
on a single node machine with 52 cores (26 dual-socket Intel Xeon Gold 6230R CPU) and 755 GB of main memory. We parallelize both versions to use the same number of cores, meaning that we cannot leverage the full multi-layer parallel scheme of our implementation, thus factorizing $\Q_{\vec{x}}$ and $\Q_{\vec{x} \vert \y}$ sequentially. Both algorithms utilize PARDISO as an underlying solver for the computational kernel operations.
We observe that for the smallest test cases they have almost identical runtimes, with INLA\textsubscript{PARDISO} exhibiting more consistent scaling properties as the model parameter dimensions grow. 
The largest test case took less than 2 hours for INLA\textsubscript{PARDISO} and just over 3.5 hours for R-INLA. 
In the following we will see that our methods INLA\textsubscript{PARDISO} and INLA\textsubscript{BTA} scale well on distributed architectures (with INLA\textsubscript{BTA} outperforming INLA\textsubscript{PARDISO} for all larger models), thus making it possible to run much larger models than the ones employed here within shorter runtimes. 

\sidecaptionvpos{figure}{c}
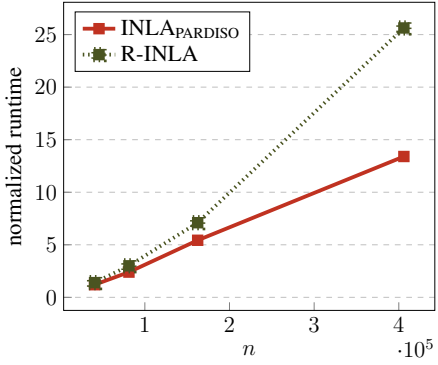
\begin{SCfigure}[1][t]
	\begin{tikzpicture}[scale=0.72, font=\large]
		\begin{axis}
			[legend pos={north west}, 
			xlabel = {$n$},
			ylabel = {normalized runtime},
			ylabel near ticks,
			ymajorgrids=true,
			grid style=dashed,
			xtick pos=left,
			ytick pos=left,
			every axis plot/.append style={thick},
			legend cell align={left}]
			
			\pgfplotstableread{benchmarks/temporal_scaling_PARDISO_ns812_nb6_9_0_4.txt}\loadedtable
			\addplot [mark=square*, mark size=2pt, myred, line width=2pt] table[col sep=space, x="n", y="timePerFn"] {\loadedtable};
			
			\pgfplotstableread{benchmarks/temporal_scaling_RINLA_ns812_nb6_9_4.txt}\loadedtable
			\addplot [mark=square*, mark size=3pt, dotted,  line width=2pt, armygreen] table[col sep=space, x="n", y="timePerFn"] {\loadedtable};			
			
			\legend{INLA\textsubscript{PARDISO},R-INLA}
			
		\end{axis}
	\end{tikzpicture}
	\hspace{15pt}
	\caption{The normalized runtime (runtime / (total \# of function evaluations)) to account for varying number of BFGS iterations until convergence, INLA\textsubscript{PARDISO} and R-INLA using the same number of cores.
		The spatial grid size is fixed to $n_s =812$ nodes with increasing number of time steps $n_t = 50, 100, 200, 500$ leading to matrix dimensions of $n \leq 4.1 \cdot 10^5$ as shown on the x-axis.}
	\label{fig:single_node_performance}
\end{SCfigure}

\subsubsection{Kernel Operations and Performance}
\label{Sec:KernelOps}

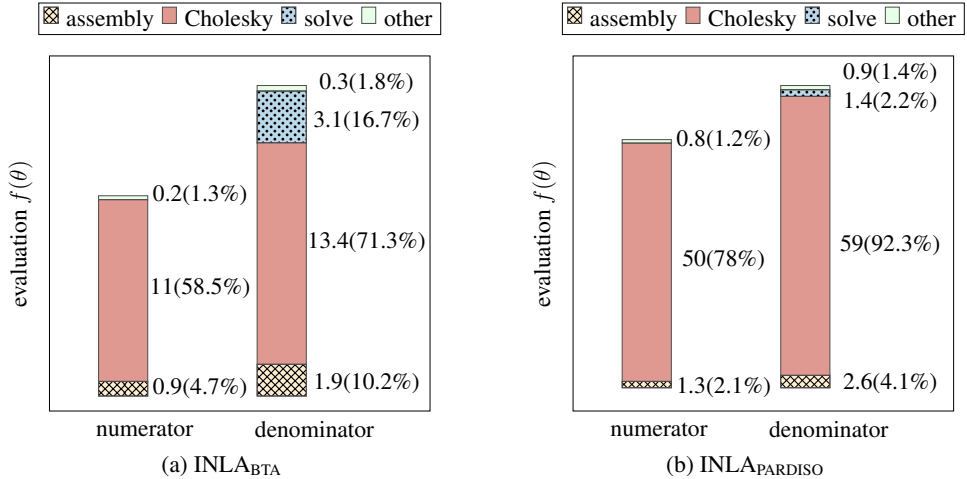
\begin{figure}[t]
	\centering
	\subfloat[INLA\textsubscript{BTA}]
	{
		\begin{tikzpicture}[scale=0.62,  font=\Large]
			\pgfplotsset{
				every axis legend/.append style={ at={(0.5,1.15)}, anchor=north,legend columns = 4, column sep=0.1cm}}
			\begin{axis}
				[x=3.4cm, 
				width=.7\linewidth,
				ybar stacked, enlarge x limits=0.7, 
				ytick style={draw=none}, 
				xtick style={draw=none}, 
				yticklabels={,,}, 
				scaled y ticks=false, 
				xtick pos=bottom,
				ybar=-40, 
				ylabel={evaluation $f(\mathbf{\theta})$}, 
				compat=1.3, 
				symbolic x coords = {numerator, denominator},
				xticklabel style = {xshift=-1.5cm, yshift=-7},
				xtick={numerator, denominator}, 
				x tick label style={anchor=west, },
				nodes near coords align={vertical}, bar width=30,
				legend image code/.code={
					\draw [#1] (0cm,-0.1cm) rectangle (0.3cm,0.25cm); },
				]
				
				]

				\node[draw=none] at (25,-20)   {0.9(4.7\%)};  
				\node[draw=none] at (25,300)   {11(58.5\%)};  
				\node[draw=none] at (25,600)   {0.2(1.3\%)};    
				
				\node[draw=none] at (130,-5) {1.9(10.2\%)}; 
				\node[draw=none] at (130,450) {13.4(71.3\%)}; 
				\node[draw=none] at (130,840) {3.1(16.7\%)}; 
				\node[draw=none] at (130,960) {0.3(1.8\%)}; 
				
				\addplot+[ybar, bar shift=-0.8cm, fill=mybeige1!90,draw=black!70, postaction={pattern=crosshatch}] plot coordinates {(numerator,4.7)(denominator,10.2)};
				\addplot+[ybar, bar shift=-0.8cm, fill=myred!50,draw=black!70] plot coordinates {(numerator,58.5)(denominator,71.3)};
				\addplot+[ybar, bar shift=-0.8cm, fill=myblue!30,draw=black!70, postaction={pattern=crosshatch dots}] plot coordinates {(numerator,0)(denominator,16.7)};
				\addplot+[ybar, bar shift=-0.8cm, fill=green!10,draw=black!70] plot coordinates {(numerator,1.3)(denominator,1.8)};
				
				\legend{assembly, Cholesky, solve, other, legend pos=north west}
			\end{axis}
		\end{tikzpicture}	
	}
	\hspace{25pt}
	\subfloat[INLA\textsubscript{PARDISO}]{
		\begin{tikzpicture}[scale=0.62,  font=\Large]
			\pgfplotsset{
				every axis legend/.append style={ at={(0.5,1.15)}, anchor=north,legend columns = 4, column sep=0.1cm}}
			\begin{axis}
				[x=3.4cm,
				width=.7\linewidth,
				ybar stacked, enlarge x limits=0.7, 
				ytick style={draw=none}, 
				xtick style={draw=none}, 
				yticklabels={,,}, 
				scaled y ticks=false, 
				xtick pos=bottom,
				ybar=-40, 
				ylabel={evaluation $f(\mathbf{\theta})$}, 
				compat=1.3, 
				symbolic x coords = {numerator, denominator},
				xticklabel style = {xshift=-1.7cm, yshift=-7},
				xtick={numerator, denominator}, 
				x tick label style={anchor=west, },
				nodes near coords align={vertical}, bar width=30,
				legend image code/.code={
					\draw [#1] (0cm,-0.1cm) rectangle (0.3cm,0.25cm); },
				]
				
				]
				\node[draw=none] at (25,-20)  {1.3(2.1\%)};  
				\node[draw=none] at (25,400) {50(78\%)}; 
				\node[draw=none] at (25,800) {0.8(1.2\%)};
				
				\node[draw=none] at (130,-5)   {2.6(4.1\%)}; 
				\node[draw=none] at (130,450) {59(92.3\%)}; 
				\node[draw=none] at (130,925)  {1.4(2.2\%)};  
				\node[draw=none] at (130,1020) {0.9(1.4\%)};   
				
				\addplot+[ybar, bar shift=-0.8cm, fill=mybeige1!90,draw=black!70, postaction={pattern=crosshatch}] plot coordinates {(numerator,2.1)(denominator,4.1)};
				\addplot+[ybar, bar shift=-0.8cm, fill=myred!50,draw=black!70] plot coordinates {(numerator,78.8)(denominator,92.3)};
				\addplot+[ybar, bar shift=-0.8cm, fill=myblue!30,draw=black!70, postaction={pattern=crosshatch dots}] plot coordinates {(numerator,0)(denominator,2.2)};
				\addplot+[ybar, bar shift=-0.8cm, fill=green!10,draw=black!70] plot coordinates {(numerator,1.2)(denominator,1.4)};
				
				\legend{assembly, Cholesky, solve, other, legend pos=north west}
			\end{axis}
		\end{tikzpicture}		
	}
	\caption{
		Computational kernel operations of $f(\th)$ for the Base Case, see Table~\ref{tab:syn_base_case}, 
		and split in the parallel evaluation of numerator and denominator, see Eq.~(\ref{eq:post_theta_approx}), using (a) INLA\textsubscript{BTA} and (b) INLA\textsubscript{PARDISO}.
		For each solver type max(runtime(numerator), runtime(denominator)) denotes $100\%$, respectively. The size of the bars is relative to this maximum. The first number next to the bar denotes the respective absolute runtime in seconds for each task, the second number denotes the respective percentage.
		For both solvers the Cholesky decompositions of $\Q_{\x}$ (numerator) and $\Q_{\x \vert \y}$ (denominator) dominate the runtimes. INLA\textsubscript{BTA} outperforms  INLA\textsubscript{PARDISO} by a factor of 3.5.
	}
	\label{fig:timespent_f_eval}
\end{figure}

In Fig.~\ref{fig:timespent_f_eval} we show the different contributions to the overall runtime for a single function evaluation of $f(\th)$, for INLA\textsubscript{BTA} and  INLA\textsubscript{PARDISO}, respectively, using the Base Case from Table~\ref{tab:syn_base_case}. 
The evaluation of $f$ is split in numerator, which entails the factorization of $\Q_{\x}$ and denominator, which entails the factorization of $\Q_{\x \vert \y}$, and are executed in parallel.
Therefore the overall runtime of a function evaluation of $f$ is determined by the maximum between the two and corresponds to $100\%$ for each solver, respectively. 
The majority of the runtime for both solvers is made up by the Cholesky factorizations.
They are performed on separate GPUs or nodes, for  INLA\textsubscript{BTA} and  INLA\textsubscript{PARDISO}, respectively.
The sparse matrix assembly for $\Q_{\x \vert \y}$ takes longer as it includes the dense arrowhead structure.
All other operations like the evaluation of the prior of the hyperparameters and the likelihood take up a negligible amount of time with a contributions of a few percent.

\begin{figure}[t]
	\centering
	\subfloat[BTA Solver]{
		\begin{tikzpicture}[scale=0.8, x=1.0cm,y=1.0cm,scale=1,baseline={(0,0)}]
			\begin{axis}[
				xticklabels={A100, A100},
				xtick=data,
				ylabel={Gflop/s},
				major x tick style=transparent,
				ybar= \pgflinewidth,
				ymax=5200,
				ymin=0,
				x tick label style={rotate=0, anchor=center},
				xticklabel style={yshift=-2mm}, 
				ymajorgrids=true,
				grid style=dotted,
				nodes near coords,
				scale only axis,
				point meta=explicit symbolic,
				enlarge x limits = {abs=1},
				cycle list={
					draw=blue,thick,fill=blue,fill opacity=0.6,nodes near coords style={blue!60}\\
					draw=orange,thick,fill=orange,fill opacity=0.6,nodes near coords style={orange}\\
				},
				legend columns=-1,
				legend pos=north east,
				height=0.3\textwidth,
				width=0.2\textwidth,
				extra x ticks={1.1, 2.1},
				extra x tick labels={$\mathbf{Q}_{\mathbf{x}}$,$\mathbf{Q}_{\mathbf{x} \vert \mathbf{y}}$},
				extra x tick style={yshift=-15pt},
				]
				
				\addplot[armygreen!80, fill]  coordinates {
					(1,5040)
					(2,4100)};
			
		\end{axis}
		\label{fig:gflops_gpu}
	\end{tikzpicture}
}
\hspace{0.5cm}
\subfloat[PARDISO]{
	\begin{tikzpicture}[scale=0.8, x=1.0cm,y=1.0cm,scale=1,baseline={(0,0)}]
		\begin{axis}[
			xticklabels={1, 2, 4, 8, 16,32, 64, 1, 2, 4, 8, 16,32, 64},
			xtick=data,
			ylabel={Gflop/s},
			major x tick style=transparent,
			ybar= \pgflinewidth,
			ymax=450,
			x tick label style={rotate=0, anchor=center},
			xticklabel style={yshift=-2mm}, 
			ymajorgrids=true,
			grid style=dotted,
			nodes near coords,
			scale only axis,
			point meta=explicit symbolic,
			enlarge x limits = {abs=1},
			cycle list={
				draw=blue,thick,fill=blue,fill opacity=0.6,nodes near coords style={blue!60}\\
				draw=orange,thick,fill=orange,fill opacity=0.6,nodes near coords style={orange}\\
			},
			legend columns=-1,
			legend pos=north east,
			height=0.3\textwidth,
			width=0.65\textwidth,
			extra x ticks={4.5, 13},
			extra x tick labels={$\mathbf{Q}_{\mathbf{x}}$,$\mathbf{Q}_{\mathbf{x} \vert \mathbf{y}}$},
			extra x tick style={yshift=-15pt},
			]
			
			\addplot[armygreen!80, fill]  coordinates {
				(1,27.5)
				(2,54.2 )
				(3,96)
				(4,185)
				(5, 345)
				(6,436)
				(7, 432)
				(9, 21.6)
				(10,41 )
				(11,77)
				(12,142)
				(13, 261)
				(14,358)
				(15, 354)};
	\end{axis}
	\label{fig:gflops_pardiso}
\end{tikzpicture}
} 
\caption{\textbf{Left-Panel:} BTA Solver. Gflop/s during the numerical factorization of $\Q_{\vec{x}}$ and $\Q_{\x \vert \y}$, respectively, using an A100 GPU. \textbf{Right-Panel:} PARDISO. Gflop/s relative to the number of cores during the numerical factorization of $\Q_{\vec{x}}$ and $\Q_{\x \vert \y}$. It can be seen that using 64 cores no longer increases Gflop/s performance. The matrices  $\Q_{\vec{x}}$ and $\Q_{\x \vert \y}$ stem from the Base Case.}
\label{fig:glfops}
\end{figure}
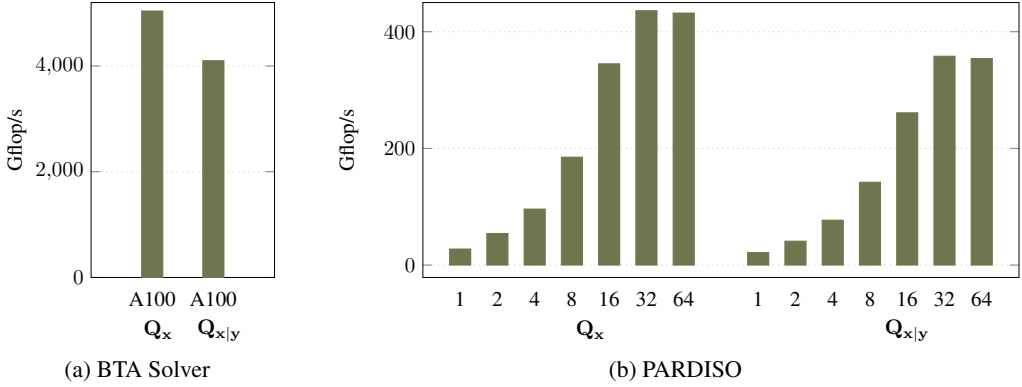

Fig.~\ref{fig:glfops} shows the achieved GFlop/s of the numerical factorization of $\Q_{\vec{x}}$ and $\Q_{\x \vert \y}$, respectively for the BTA solver and PARDISO. As expected the former shows a much higher GFlop/s performance than the latter as it operates entirely on dense submatrices as compared to the sparse structures.
Additionally we can see how the number of floating point operations per second increases with the number of cores used by PARDISO up to 32 cores, where it peaks. 

\subsubsection{Strong Scaling}
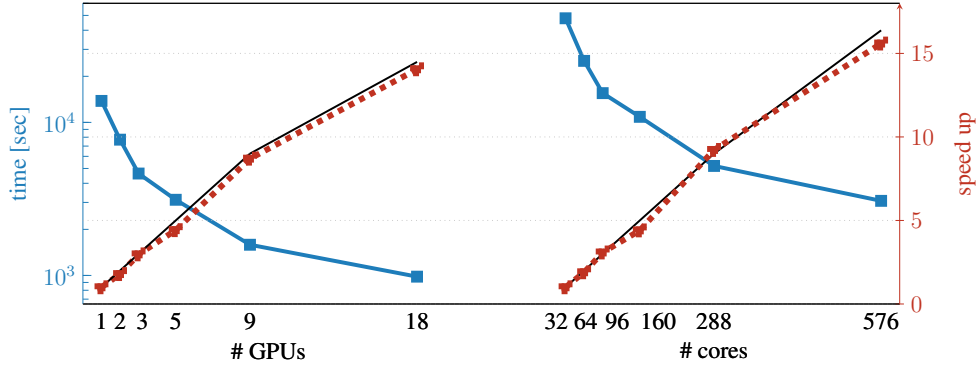
\begin{figure}[t]
\centering
\begin{tikzpicture}[scale=0.75, x=1.0cm,y=1cm,baseline={(0,0)}, font=\large]
\begin{axis}[
	xticklabels={1,2,3, 5, 9,18, 32, 64, 96, 160, 288, 576},
	xtick={1,2,3.2,5, 9,18 ,25.5, 27.1, 28.8, 31, 34, 43},
	ylabel={time [sec]},
	ymode=log,	
	ylabel style={myblue}, 
	axis y line*=left,
	major x tick style=transparent,
	ymax=60000,
	x tick label style={rotate=0, anchor=center},
	xticklabel style={yshift=-3mm}, 
	ylabel near ticks, yticklabel pos=right,
	ylabel style={yshift=-0.0cm, color=myblue}, 
	y tick style= {myblue},
	y tick label style= {myblue},
	y axis line style = {myblue},
	ymajorgrids=false,
	grid style=dotted,
	nodes near coords,
	scale only axis,
	point meta=explicit symbolic,
	enlarge x limits = {abs=1},
	height=0.35\textwidth,
	width=0.95\textwidth,
	extra x ticks={10, 34},
	extra x tick labels={\# GPUs, \# cores},
	extra x tick style={yshift=-15pt},
	legend cell align={left},
	]
	
	\addplot[myblue, mark=square*, line width=2 pt]  coordinates {
		(1,13815)
		(2, 7712)
		(3, 4633)
		(5, 3122)
		(9, 1587)
		(18, 982)
	};
	
	\addplot[myblue, mark=square*, line width=2 pt] coordinates{			%
		(26,47888.328 )  			 
		(27, 25230.602)			 
		(28, 15531.086)	         
		(30, 10872.032 )           
		(34, 5193.517)             
		(43,3071.616)             
	};

\end{axis}
\begin{axis}[
	xticklabels={1,2,3, 5, 9,18, 32, 64, 96, 160, 288, 576},
	xtick={1,2,3.2,5, 9,18 ,25.5, 27.1, 28.8, 31, 34, 43},
	ylabel={speed up},
	axis y line=right,
	major x tick style=transparent,
	ymin = 0,
	ymax=18,
	x tick label style={rotate=0, anchor=center},
	xticklabel style={yshift=-3mm}, 
	ylabel style={yshift=+0cm, color=myred}, 
	y tick style= {myred},
	y tick label style= {myred},
	y axis line style = {myred},
	ymajorgrids=true,
	grid style=dotted,
	nodes near coords,
	scale only axis,
	point meta=explicit symbolic,
	enlarge x limits = {abs=1},
	legend columns=-1,
	legend pos=north east,
	height=0.35\textwidth,
	width=0.95\textwidth,
	extra x ticks={10, 34},
	extra x tick labels={\# GPUs, \# cores},
	extra x tick style={yshift=-15pt},
	]
	
	\addplot[domain=26:34, black,  line width=1]{x-25};
	\addplot[domain=34:43, black,  line width=1]{0.82*x-25+6.12};
	
	\addplot[myred, dashed, mark=square*, line width=3pt]  coordinates {
		(1,1)
		(2, 1.79)
		(3, 2.97)
		(5, 4.42)
		(9, 8.7)
		(18, 14.05)
	};
	
	\addplot[domain=1:9, black,  line width=1]{x};
	\addplot[domain=9:18, black,  line width=1]{0.61*x+3.51};
	
	\addplot[myred, dashed, mark=square*, line width=3pt] coordinates{			%
		(26,1)
		(27, 1.898026)
		(28,  3.083386 )
		(30, 4.404727  )
		(34, 9.220790 )
		(43, 15.590595)
	};

\end{axis}
\end{tikzpicture}	
\caption{Strong scaling plot of fitting the Base Case model. The blue line indicates the runtime in seconds using different numbers of GPUs or cores, respectively. The red line indicates the speed up over the respective base version (1 GPU/32 cores). The black dashed line represents the ideally achievable speed up. \textbf{Left-Panel:} INLA\textsubscript{BTA}. \textbf{Right-Panel:} INLA\textsubscript{PARDISO}.}
\label{fig:parallel_scaling}
\end{figure}

In this section we discuss performance results with respect to increasing numbers of GPUs/cores. 
The majority of the overall runtime of the algorithm is spent on solving the optimization problem described in Sect.~\ref{sec:INLA_methodology}, Step~\ref{enum:INLA_step1}. 
In each iteration of the BFGS solver $2 \,d(\th) + 1$ parallel function evaluations are required to compute $f(\th^k)$ and $\nabla f(\th^k)$.
One can observe in Fig.~\ref{fig:parallel_scaling} that our algorithm actually exhibits ideal scaling for up to 9 processes (dim$(\th) = 4$) as expected.

As previously discussed, further parallelism is introduced by simultaneously factorizing $\Q_{\x}$ and $\Q_{\x | \y}$.
Since $\Q_{\x}$ has less nonzero entries and a simpler sparsity structure, its factorization takes less time than of $\Q_{\x | \y}$.
Therefore, the theoretical maximum speedup going from sequentially factorizing these two matrices to parallel factorizations is less than a factor 2, but rather around a factor 1.6 for  INLA\textsubscript{BTA} and 1.8 for  INLA\textsubscript{PARDISO}, see Fig.~\ref{fig:timespent_f_eval} for details. 
This also becomes apparent in our numerical experiments. When parallelizing the Cholesky decompositions within each function evaluation on top of the parallel function evaluations, we can observe a further speed up that corresponds approximately to the theoretically achievable factor of $1.6$ and $1.8$ as respectively, indicated by the black lines.
In terms of wall-clock time we can see that INLA\textsubscript{BTA} outperforms PARDISO by roughly a factor 3-4, independently of the number of processes. 
To fit the Base Case model, INLA\textsubscript{BTA} requires just under $1 \cdot 10^3$ seconds, i.e. about 16 minutes, when using 18 GPUs. INLA\textsubscript{PARDISO} requires just over 50 minutes when using 576 cores.

\subsubsection{Temporal and Spatial Scaling}

\begin{figure}[t]
	\centering
\subfloat[Temporal Scaling]{
\begin{tikzpicture}[scale=0.73, font=\Large]
	\begin{axis}
		[legend pos={north west}, 
		xlabel = {$n$},
		ylabel = {normalized runtime},
		xmax=2.24*10^6,
		ylabel near ticks,
		ymajorgrids=true,
		grid style=dashed,
		xtick pos=left,
		ytick pos=left,
		every axis plot/.append style={thick},
		legend cell align={left},]
		
		\pgfplotstableread{benchmarks/median_temporal_scaling_results_PARDISO_ns4002_nb6_18_0_32_relBFGSdelta1e-7.txt}\loadedtable
		\addplot [mark=square*, myred,  line width=2pt] table[col sep=space, x="n", y="timePerFn"] {\loadedtable};
		
		\pgfplotstableread{benchmarks/median_temporal_scaling_results_RGF_ns4002_nb6_9_2_1_properPin_relBFGSdelta1e-7.txt}\loadedtable
		\addplot [mark=star, mark size=3pt, myblue, line width=3pt, dashed] table[col sep=space, x="n", y="timePerFn"] {\loadedtable};
		
		\addplot[domain=200106:2400000, black,  line width=1]{ 2.212e-06*x-1.327e-05};

		\legend{$\text{INLA}_{\text{PARDISO}}$,$\text{INLA}_{\text{BTA}}$, $O(n_t)$}
		
	\end{axis}
	\label{fig:temporal_scaling}
\end{tikzpicture}
}
\hspace{20pt}
\subfloat[Spatial Scaling]{
\begin{tikzpicture}[scale=0.73, font=\Large]
	\begin{axis}
		[legend pos={north west}, 
		xlabel = {$n$},
		ylabel = {normalized runtime},
		ylabel near ticks,
		ymajorgrids=true,
		grid style=dashed,
		xtick pos=left,
		ytick pos=left,
		every axis plot/.append style={thick},
		legend cell align={left},]

		\pgfplotstableread{benchmarks/median_spatial_scaling_results_PARDISO_nt30_nb6_9_2_32_relBFGSdelta1e-7.txt}\loadedtable
		\addplot [mark=square*, myred,  line width=2pt] table[col sep=space, x="n", y="timePerFn"]   {\loadedtable};
		
		\pgfplotstableread{benchmarks/median_spatial_scaling_results_RGF_nt30_nb6_9_2_1_properPin_relBFGSdelta1e-7.txt}\loadedtable
		\addplot [mark=star, mark size=3pt, myblue, line width=3pt, dashed] table[col sep=space, x="n", y="timePerFn"] {\loadedtable};
													
		\legend{$\text{INLA}_{\text{PARDISO}}$,$\text{INLA}_{\text{BTA}}$}
		
	\end{axis}
	\label{fig:spatial_scaling}
\end{tikzpicture}%
}

\caption{
The normalized runtime (runtime / (total \# of function evaluations)) over the precision matrix dimension $n$. 
\textbf{Left Panel:} Fixed spatial grid size of $n_s =4002$ nodes with increasing number of time steps, corresponding to TS I-II, BC \& TS III from Table~\ref{tab:syn_base_case}. The black dotted line (coinciding with INLA\textsubscript{BTA})  indicates the linear complexity in $n_t$ as discussed in Table~\ref{tab:complexity}. \textbf{Right Panel:} Fixed temporal grid size of $n_t = 30$ nodes with increasing number of spatial nodes, corresponding to SS I-IV.}
\label{fig:temporal_spatial_scaling}
\end{figure}
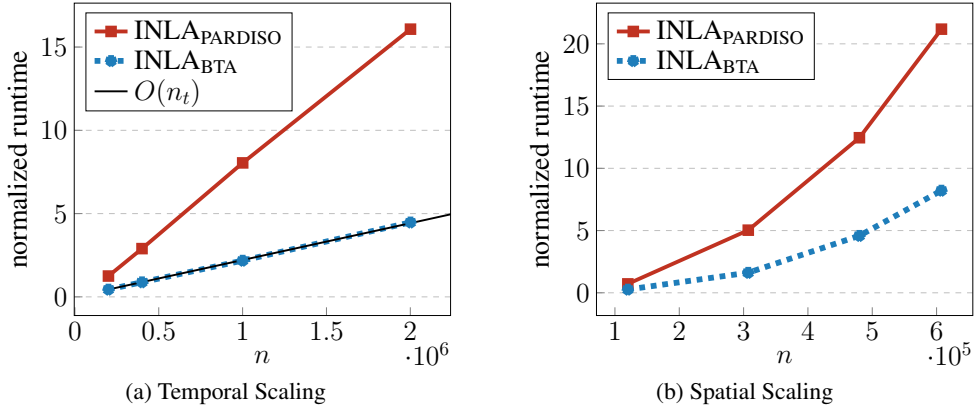

We will now discuss how our proposed methods scale with increasing numbers of spatial $n_s$ and temporal nodes  $n_t$ , respectively. 
The normalized runtimes (total runtime divided by total number of function evaluations) are shown with respect to the matrix size in Fig.~\ref{fig:temporal_spatial_scaling}. 
We note that for both scaling studies, the INLA\textsubscript{BTA} outperforms INLA\textsubscript{PARDISO} by roughly a factor of 3-4 for larger problems in terms of total runtime.
The analysis in Sect.~\ref{sec:complexity} showed that the BTA solver has linear complexity in time for a single matrix factorization or selected inversion. As these are the dominating subroutines of INLA\textsubscript{BTA}, we would ideally expect this property to be inherited.
We can see that this holds true almost perfectly as indicated in the graph. 
PARDISO's complexity with respect to temporal nodes $n_t$ grows with a larger factor, see Table~\ref{tab:complexity}. This is in line with the observed temporal scaling behaviour of INLA\textsubscript{PARDISO}. 
In the spatial scaling case, the complexity for both solvers is superlinear which is consistent with the results shown in Fig.~\ref{fig:spatial_scaling}.
Due to the sparsity preserving SPDE approach our method scales independently of the number of observations, i.e. increasing the number of observations does not increase the runtime, see Sect.~\ref{sec:INLA_methodology} for details.

\begin{figure}[t]
\centering
{\label{fig:measurement_stations}
\includegraphics[width=.45\textwidth]{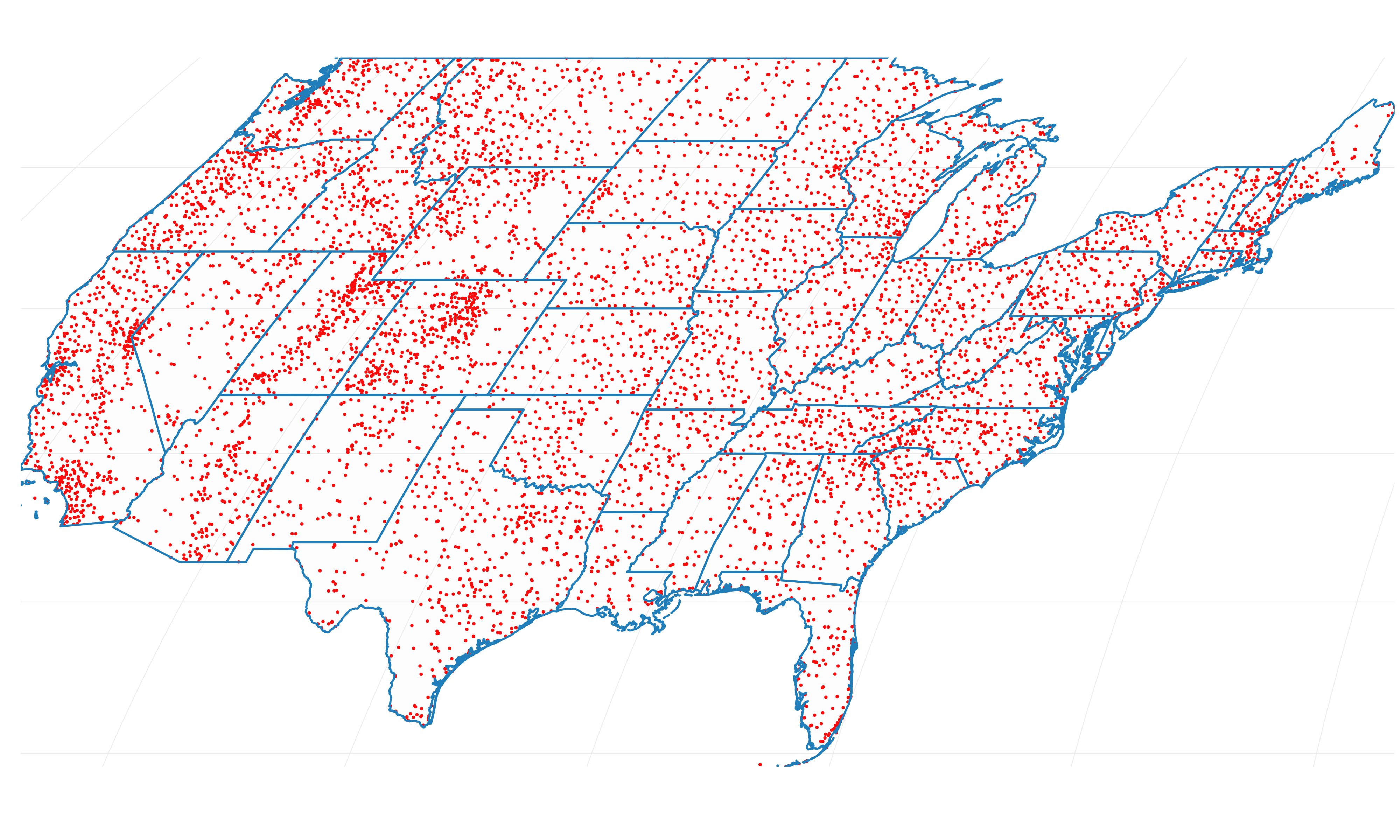}}
{\label{fig:spatial_mesh}
\includegraphics[width=.45\textwidth]{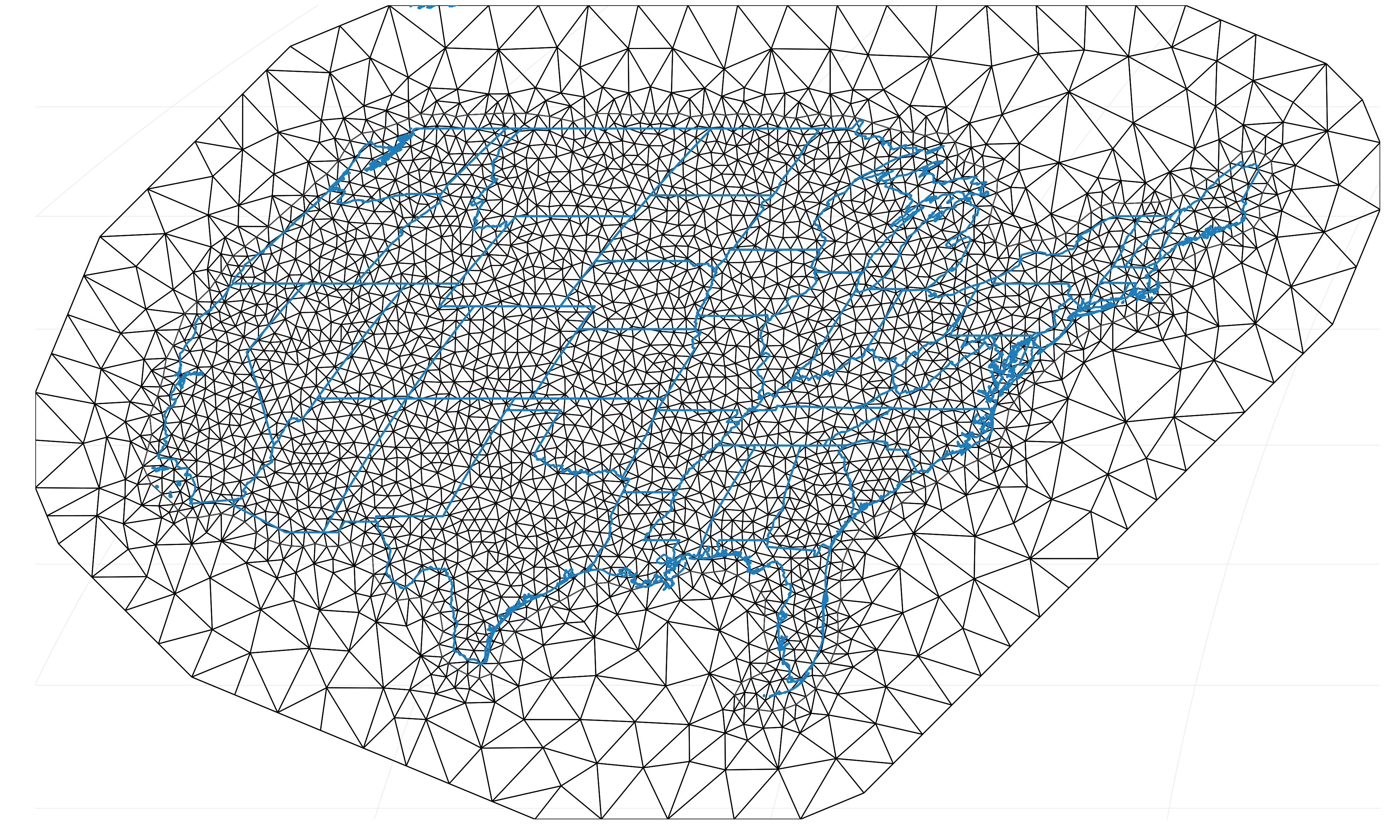}}
\caption{\textbf{Left-Panel:} The stations where measurements were collected. Not every station collects measurements at every time step.
\textbf{Right-Panel:} A map of the different US states is outlined in blue. The employed spatial disrectization is shown in grey including the coarser peripheral mesh with a total of $n_s = 2865$ mesh nodes.}
\end{figure}
\begin{figure}[t]
\centering
{\label{fig:spatially_aggregated_data}
\includegraphics[width=.53\textwidth]{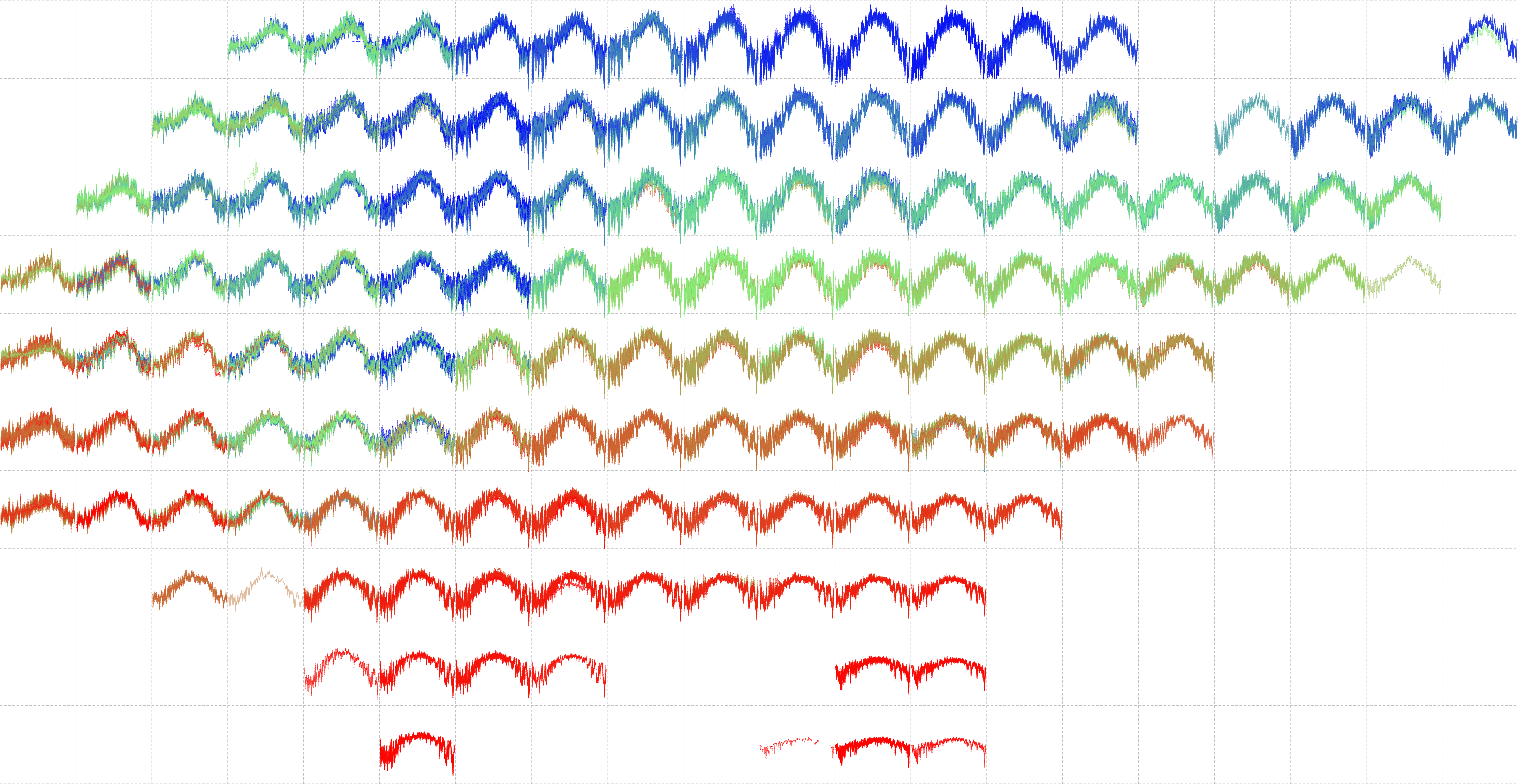}}
\caption{ The data shown as time series grouped around its locations and
colored as function of its average from colder (blue) to warmer (red).}
\end{figure}

\begin{figure}[t]
\centering
\label{fig:u_st1}
\includegraphics[width=.9\linewidth]{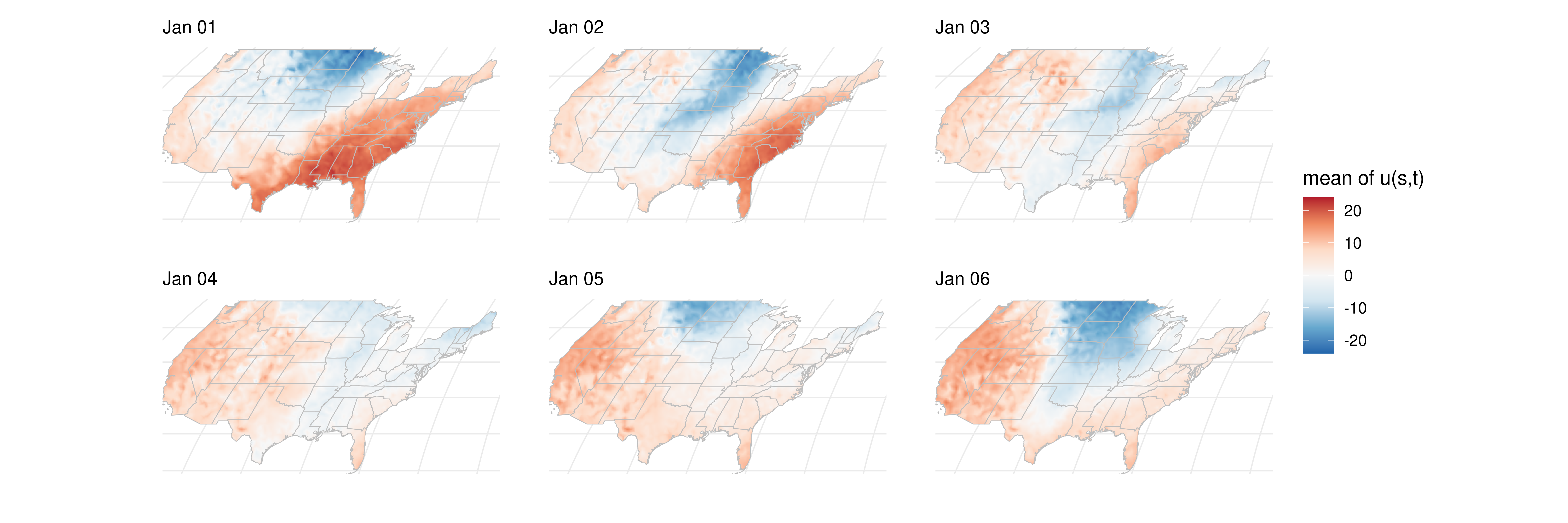}
\includegraphics[width=.9\linewidth]{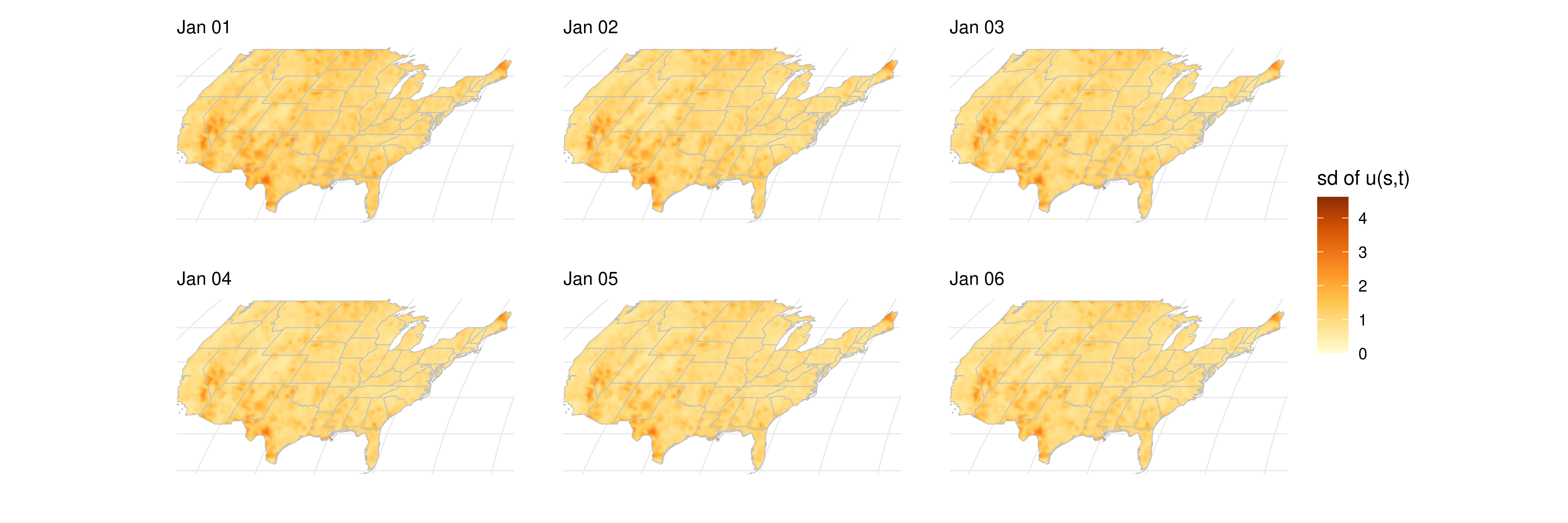}
\caption{The posterior mean and std. dev. for the $u(s,t)$ field for the first 6 days, that is Jan 1-6.}
\end{figure}
\begin{figure}[h!]
\centering
\label{fig:u_st}
\includegraphics[width=.9\linewidth]{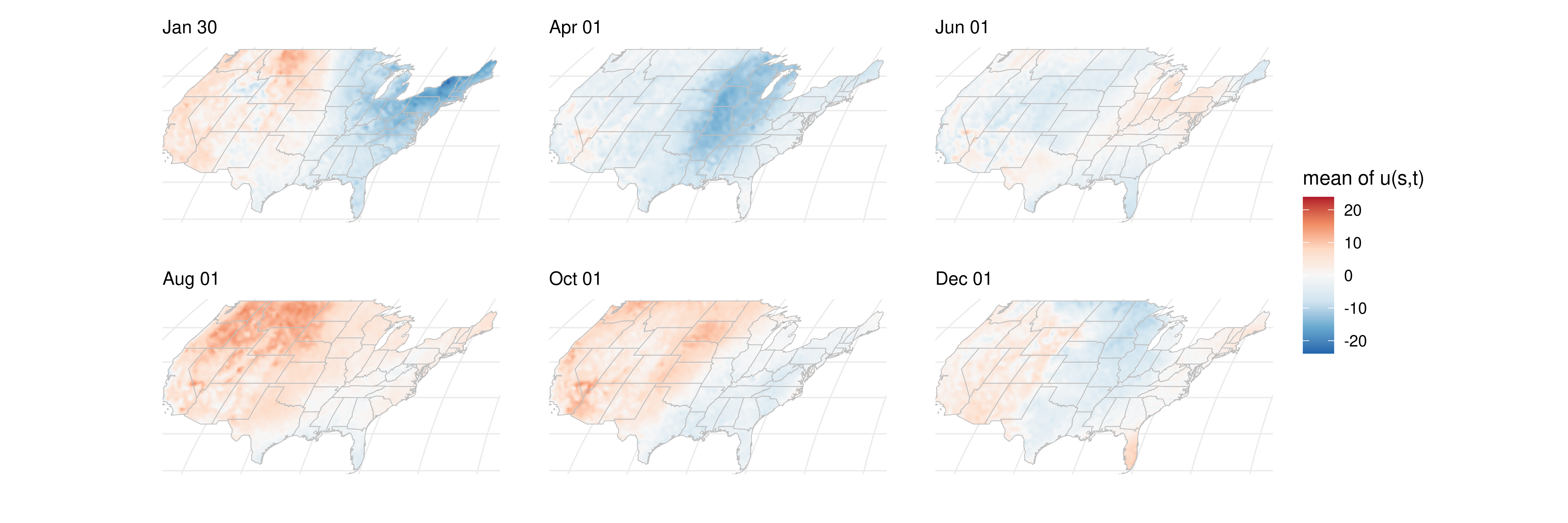}
\includegraphics[width=.9\linewidth]{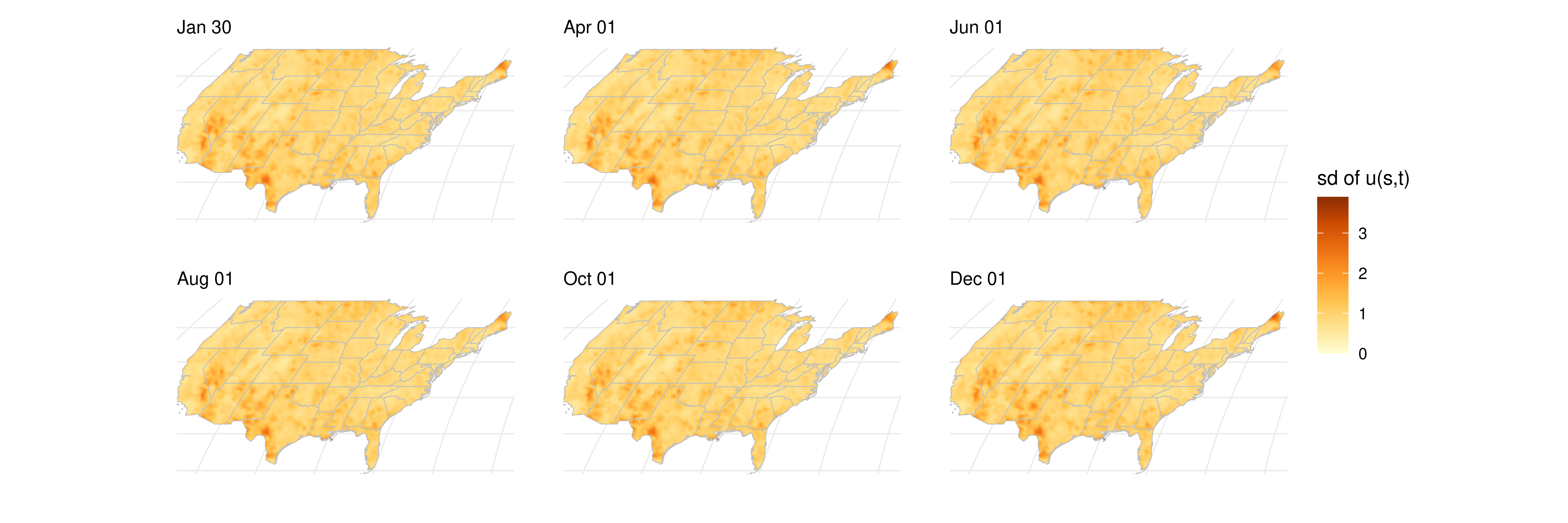}
\caption{The posterior mean and standard deviation for the $u(s,t)$ field at selected times.}
\end{figure}

\subsection{Climate Dataset Application}

In this section, we utilize our proposed method to model the daily average air temperatures over the United States for the entire year of 2022. 
Temperature measurements from different meteorological stations throughout the country are supplied by the National Centers of Environmental Information and publicly available~\cite{menne2012global}. 
We define the linear predictor $\eta(s,t)$ over the spatial-temporal domain as
\begin{equation}
\eta(s,t) = \beta_0 + \beta_1 \, \text{elevation}(s) + \beta_2 \,  \text{z}(t) + \beta_3 \, \text{north}(s) + u(s,t),
\end{equation}
where $\beta_0$ represents an overall temperature mean and
$\beta_1$ is the effect of elevation at a location $s$.
The term $z(t) = \textrm{sin}(\pi t/365)$ describes a normalized seasonal trend and
$\beta_2$ is the effect of this term.
The variable $\beta_3$ represents the effect of the projected latitude with units in kilometers from the southeast data location, 
and $u(s,t)$ is the spatial-temporal field.
The unknown posterior parameter distributions in the linear predictor 
are the fixed effects $\vec{\beta} = (\beta_0, ..., \beta_3)^T$, as well as the random field $u(s,t)$.
The spatial-temporal domain is discretized using first order finite elements, whose basis functions are defined on the spatial-temporal mesh nodes, as detailed in \cite{lindgren2022diffusion}. 
For simplicity we keep the same spatial discretization at each time step. 
We embed the contiguous US territory, i.e. omitting Hawaii and Alaska, in a two-dimensional convex domain that is discretized using a Delauney triangulation and add a coarser peripheral mesh to buffer boundary effects.
The different measurement locations and the spatial mesh
are shown in Fig.~\ref{fig:measurement_stations}.  
The temporal mesh consists of $n_t = 365$ nodes, one for each day. It is worth noting that not all stations have measurements at every time step.
There are about $2.5$ million observations available throughout the entire year, see Fig.~\ref{fig:spatially_aggregated_data}. 
The arising latent parameter space has a dimension of over $1$ million, see Table~\ref{tab:temp_dataset} for details. 
\begin{table}[h!]
	\centering
	\begin{tabular}{r|r|r|r|r}
		$n_s$& $n_t$ & $n_b$& $n_o$ & $n$  \\ \hline \hline
		2865 & 365 & 4 & 2 472 561 & 1 045 729 
		\vspace{6pt}
	\end{tabular}
	\caption{Model parameter dimensions for the air temperature dataset, following the notation of Table~\ref{tab:syn_base_case}.}
	\vspace{-5mm}
	\label{tab:temp_dataset}
\end{table}

\begin{figure}
\begin{minipage}{.55\linewidth}
\centering
\begin{tabular}{l|r|r|r|r}
	& $\beta_0$ &  $\beta_1$ & $\beta_2$ & $\beta_3$  \\ \hline \hline
	est. mean & 10.58 & -3.84 & 25.70  & -7.37  \\ \hline 
	est. sd & 0.35 & 0.006 & 0.34 & 0.16 
	\vspace{4pt}
\end{tabular}
\end{minipage}
\begin{minipage}{.37\linewidth}
\centering
\begin{tikzpicture}[scale=0.44, font=\Large]
	\begin{axis}  
		[  
		ybar,  
		ymax=4,
		ymin=0,
		ylabel style = {xshift=0cm, yshift=+0.5cm},
		ylabel={normalized runtime}, 
		symbolic x coords={INLA-PARDISO, INLA-BTA}, 
		xtick=data,  
		nodes near coords, 
		nodes near coords align={vertical},  
		bar width=40,
		enlarge x limits=0.5,
		xticklabel style={text height=2ex, rotate=60, anchor = north},
		xticklabel style={xshift=-9mm, yshift=-0mm, align=center,text width=23mm}, 
		]  
		
		\addplot[armygreen!80, fill]  coordinates {
			(INLA-PARDISO,3.3 )
			(INLA-BTA,2.05)
		};
		
	\end{axis}  
\end{tikzpicture}
\end{minipage}
\caption{\textbf{Left-Panel:} Estimated posterior fixed effects. \textbf{Right-Panel: } Normalized runtime in seconds per function evaluation for INLA\textsubscript{PARDISO} and INLA\textsubscript{BTA}.} 
\label{fig:temp_results_fixed_effects_runtime}
\end{figure}
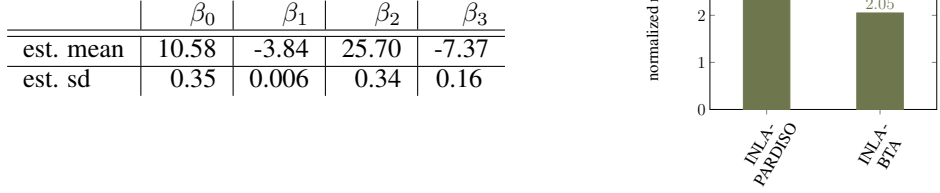

The first inference step is to estimate the hyperparameter distributions of $\th$ as described in 
Sect.~\ref{sec:SPDE_approach}, which parametrize the 
noise term and the spatio-temporal field. 
From this the posterior mean and standard deviation for the fixed effects $\vec{\beta}$ are derived and presented in Fig.~\ref{fig:temp_results_fixed_effects_runtime}.  
They give rise to the following model interpretation. 
The estimated temperature in 2022 was 10.6 degrees Celcius, in the beginning and at the end of the year,
at the southeast station location at sea level. 
It decreased 3.8 degrees per kilometer of elevation.
On average the temperature increased by 25.7 degrees by the middle of the year,
and decreases 7.37 degrees per thousand kilometers going north from the most southeastern point.
While general trends are captured by the fixed effects, the spatial-temporal field $u(s,t)$ describe 
the local deviations from this overall behavior. 
In Fig.~\ref{fig:u_st1} the estimated posterior mean and standard deviation of $u(s,t)$ are shown for the 
first six days of the year 2022.
It can be seen that on January 1st it was around 10 to 20 Celcius degrees 
warmer, than the fixed effects would stipulate in the southeastern part of the domain. 
The warmer region shrinks in size over the following days, 
showing the ability of this term to indeed capture short-term variations.
We note that the standard deviation of $u(s,t)$ is lower in areas with a higher density of measurement stations and larger in areas with fewer measurement stations, see for example the northeastern part of the country as well as in some areas of Texas. 
In Fig.~\ref{fig:u_st} we show the posterior mean and standard deviation for $u(s,t)$ for selected days over the year which are apart around a period of two months. 
In these far apart in time maps we can see a different spatial pattern for the posterior mean as they are related to the weather at these days which are reasonable to be assumed uncorrelated. 
We notice that with the spatial resolution used to solve the model 
it was able to capture the local behavior of the weather.
The posterior standard deviation looks similar for each shown time steps which corresponds to the fact that most stations record observations daily.

The normalized runtimes are shown in Fig.~\ref{fig:temp_results_fixed_effects_runtime} using the most performant configuration for each version, respectively. The overall runtimes were just over 25 for INLA\textsubscript{PARDISO} and 15 minutes for INLA\textsubscript{BTA}.

\section{Conclusion}
\label{conclusion}

In this work we propose a novel scheme for large-scale Bayesian inference tasks for spatial-temporal data. 
The INLA methodology in combination with the SPDE approach provide an efficient framework for performing spatial-temporal Bayesian modeling, as they are able to use sparse representations of the underlying processes. 
Building upon this work, we used a non-separable space-time model to capture spatial-temporal phenomena.
We introduced a distributed memory approach, INLA\textsubscript{DIST}, which employs a multi-layer parallel scheme that operates within and across nodes.
It takes advantage of mutually independent function evaluations in the optimization phase, identifies parallelizable kernel operations, and exploits concurrency within each subroutine. 
The spatial-temporal discretization of the model induces block tridiagonal arrowhead sparsity structures in its precision matrices.
To handle the associated computational bottleneck operations, Cholesky decompositions, solving linear systems and selected matrix inversions, we derived algorithmic solutions that are tailored to the recurring sparsity patterns.
In particular we proposed an efficient selected block inversion routine to compute the marginal variances of the latent parameters, without having to compute the full inverse.
We put forward a GPU-accelerated implementation of these routines entirely based on dense block operations, while maintaining sparsity in the overall system.
We also included an alternative numerical solver for the computational key operations which is CPU-based and leverages a fully sparse approach.
In our performance analysis we show that INLA\textsubscript{DIST} exhibits almost ideal strong scaling up to two times the hyperparameter dimension. 
Its blocked GPU-based variant scales linearly with the number of time steps and quasilinearly for the sparse CPU-based variant. 
We utilize INLA\textsubscript{DIST} for a large-scale air temperature modeling application, using more than 1 million latent parameters and 2.5 million observations. 
We exhibit runtimes in the order of tens of minutes, outperforming existing approaches. 
INLA\textsubscript{DIST} allows for performing Bayesian inference on large-scale spatial-temporal data observed on unstructured planar grids or two-dimensional manifolds, like the sphere.

\section{Acknowledgments}
The authors gratefully acknowledge the great scientific support and HPC resources provided by the Erlangen National High Performance Computing Center (NHR@FAU) of the Friedrich-Alexander-Universität Erlangen Nürnberg (FAU) under the NHR project 286745. NHR funding is provided by federal and Bavarian state authorities. NHR@FAU hardware is partially funded by the German Research Foundation (DFG) – 440719683. 
The authors would also like to gratefully acknowledge the useful discussions and initial software support on the  selected block inversion from Prof. Mathieu Luisier. The OMEN \cite{luisier2008omen} software infrastructure was used as a starting point to derive and implement the BTA solver.

\bibliographystyle{unsrt}

\end{document}